\RequirePackage{ifpdf}
\ifpdf % We are running pdfTeX in pdf mode
\documentclass[pdftex]{sigma}
\else
\documentclass{sigma}
\fi

\renewcommand{\ll}{\left<\!\left<}
\renewcommand{\gg}{\right>\!\right>}

\newcommand{\R}{\mathbb{R}}
\newcommand{\A}{{\cal A}}
\newcommand{\B}{{\cal B}}

\newcommand{\e}{\epsilon}
\newcommand{\lag}{\mathfrak{g}}
\newcommand{\lak}{\mathfrak{K}}

\newcommand{\Or}{{\cal O}}
\def\fpd#1#2{\frac{\partial #1}{\partial #2}}
\newcommand{\F}{\mathbb{F}}

\newcommand{\ractie}{\psi}

\begin{document}

\allowdisplaybreaks

\renewcommand{\PaperNumber}{109}

\FirstPageHeading

\ShortArticleName{Routh Reduction by Stages}

\ArticleName{Routh Reduction by Stages}

\Author{Bavo LANGEROCK~$^{\dag\ddag\S}$, Tom MESTDAG~$^\dag$ and Joris VANKERSCHAVER~$^{\dag\star}$}

\AuthorNameForHeading{B.~Langerock, T.~Mestdag and J.~Vankerschaver}

\Address{$^\dag$~Department of Mathematics, Ghent University,
Krijgslaan 281, S22, B9000 Ghent, Belgium}
\EmailDD{\href{mailto:bavo.langerock@ugent.be}{bavo.langerock@ugent.be}, \href{mailto:tom.mestdag@ugent.be}{tom.mestdag@ugent.be}, \href{mailto:joris.vankerschaver@ugent.be}{joris.vankerschaver@ugent.be}}

\Address{$^\ddag$~Belgian Institute for Space Aeronomy,  Ringlaan 3, B1180 Brussels, Belgium}

\Address{$^\S$~Department of Mathematics, K.U. Leuven, Celestijnenlaan 200 B, B3001 Leuven, Belgium}

\Address{$^\star$~Department of Mathematics, University of California at San Diego,\\
\hphantom{$^\star$}~9500 Gilman Drive, San Diego CA 92093-0112, USA}

\ArticleDates{Received June 16, 2011, in f\/inal form November 22, 2011;  Published online November 29, 2011}

\Abstract{This paper deals with the Lagrangian analogue of symplectic or point reduction by stages. We develop Routh reduction as a reduction technique that  preserves the Lagrangian nature of the dynamics. To do so we heavily rely on the relation between Routh reduction and cotangent symplectic reduction. The main results in this paper are: (i)~we develop a class of so called magnetic Lagrangian systems and this class has the property that it is closed under Routh reduction; (ii)~we construct a transformation relating the magnetic Lagrangian system obtained after two subsequent Routh reductions and the magnetic Lagrangian system obtained after Routh reduction w.r.t.\ to the full symmetry group.}


\Classification{37J05; 37J15; 52D20}

{\small

\tableofcontents

}

\section{Introduction}

It is not a surprise that the bigger part of the recent literature on the geometric description of mechanical systems deals in some way or another with symmetry.
The benef\/its of exploiting symmetry are indeed self-evident: The equations of motion of a mechanical system which exhibits a symmetry  can be reduced to a new set of
equations with fewer unknowns, possibly easier to solve. In particular, the Hamiltonian description of such systems has attracted most of the attention in the literature, and the important role played by Poisson manifolds and symplectic
structures has been strongly emphasized (see e.g.~\cite{MarsdenHamRed} and references therein). Less well-known is the process of
symmetry reduction for Lagrangian systems. Much like on the Hamiltonian side, there are in fact
two dif\/ferent paths that lead to two dif\/ferent Lagrangian reduction
theories.  Roughly speaking, the
invariance of the Lagrangian leads via Noether's theorem to a set
of conserved quantities (the momenta). Whether or not one takes these conserved quantities into account in the reduction
process leads to either the Routh or the Lagrange--Poincar\'e
reduction method (see e.g.~\cite{CMR01,survey,mestcram,BC,marsdenrouth,MCPaper}).

In this paper we deal with Routh's reduction procedure. In a way, one restricts the attention to only those solutions of the system with a prescribed value of momentum. The price one has to pay is that
the new symmetry group (after restriction) is in general only a subgroup of the symmetry
group of the original Lagrangian system. In \cite{BC} it was pointed out that  Routh reduction can be interpreted as a special case of symplectic reduction, which moreover preserves the Lagrangian nature of
the system.

In the current paper we will investigate those aspects
of the theory that are related to reduction in multiple stages. In
case the symmetry group of the system has a normal subgroup, one can
indeed f\/irst perform a Routh reduction by means of this subgroup.
It is then natural to ask whether the reduced system is invariant
under some residual, as yet to be determined group, and whether a
subsequent second Routh reduction leads to results which are
equivalent to direct reduction by the full group. The answer to the same question, but for the above mentioned Lagrange--Poincar\'e
reduction theory, is by now well-known \cite{CMR01}.  It can be understood in the following sense: After each reduction step
one remains in the category of so-called `Lagrangian systems on Lie
algebroids'  and symmetry reduction can be phrased in terms of Lie
algebroid morphisms (see e.g.~\cite{survey} and the references therein). Equivalently, one may use a connection to decompose the Lie algebroid structure at each stage. The corresponding category is then the so-called category of
`Lagrange--Poincar\'e  bundles' from \cite{CMR01}, while the corresponding `Lagrange--Poincar\'e morphisms' do the reduction.

It is natural to ask whether or not such a
category of systems exists also for Routh reduction (it is mentioned as an open problem in~\cite{MarsdenHamRed}).  The candidate we propose (in Section~\ref{section3}) is the class of what we call `magnetic Lagrangian systems'. We show in Sections~\ref{section4} and~\ref{section5} that this
class has the property that it is closed under Routh reduction, i.e.\ that after each step
of a reduction in stages the reduced system remains in the class of magnetic Lagrangian
systems. As in~\cite{BC}, we heavily rely on a generalized version of cotangent symplectic reduction. In Section~\ref{section6} we apply the framework of magnetic Lagrangian systems to reduction in several stages, where the f\/irst stage consists of Routh reduction by a normal subgroup of the overall symmetry group. We end the paper with some illustrative examples in the f\/inal section.

\section{Preliminaries on Routh reduction}\label{section2}

\begin{definition} A Lagrangian system is a pair $(Q,L)$ with $Q$ a manifold and $L$ a function on~$TQ$. The manifold~$Q$ is the conf\/iguration space and $L$ is called the Lagrangian. The dimension of $Q$ is called the number of degrees of freedom of the Lagrangian system.  A Lagrangian system is of mechanical type if for arbitrary $v_q\in T_qQ$, the Lagrangian can be written as \[L(v_q) = \frac12 \ll v_q,v_q\gg_Q-V(q),\] with $\ll\cdot,\cdot\gg_Q$ a Riemannian metric on $Q$ and $V$ a function on $Q$, called the potential energy of the Lagrangian. The function $\frac12\ll v_q,v_q\gg_Q$ is called the kinetic energy.
\end{definition}

We are interested in the Euler--Lagrange equations associated to a Lagrangian system. These equations are necessary conditions for a curve $q(t):I\subset \R \to Q$ to extremize the action integral $\int_I L(q(t), \dot q(t)) dt$. In a local coordinate chart $(q^i)$  on $Q$ the Lagrangian $L$ is a function of $(q^i,v^i)$ and the Euler--Lagrange equations are
\[
\frac{d}{dt} \left(\fpd{L}{v^i} \right) - \fpd{L}{q^i} = 0, \qquad \dot q^i =v^i, \qquad i=1,\ldots, n=\dim Q.
\]
Routh reduction is a reduction technique for Lagrangian systems that are invariant under the action of a symmetry group. In classical textbooks, it is a step-by-step procedure that describes the construction of a new Lagrangian function with fewer degrees of freedom.

Before we can formulate Routh reduction, we f\/ix notations for the action of a Lie group on an arbitrary manifold $M$ and related concepts.

{\bf Actions of Lie groups and principal bundles.}  When a Lie group $G$ is given,  $\lag$ denotes the Lie algebra of $G$ and $\exp$ the exponential map from $\lag$ to $G$. The adjoint action of $G$ on $\lag$ is denoted by ${\rm Ad}$ and the coadjoint action of $G$ on $\lag^*$ by ${\rm Ad}^*$.

\begin{definition} \label{def:acties}\mbox{}
\begin{enumerate}\itemsep=0pt
\item A right action of a group $G$ on a manifold $M$ is denoted by $\Psi^M: G\times M\to M; (g,m)\mapsto \Psi^M_g(m)= mg$. Throughout this paper we only consider free and proper actions of Lie groups on manifolds.
\item  The inf\/initesimal action is given by  $\ractie^M_m:\lag\to T_mM; \xi \mapsto d/d\e|_0 (m\exp \e\xi)$.
\item For any element $\xi \in \lag$ one can consider the fundamental vector f\/ield $\xi_M$, def\/ined pointwise as $\xi_M(m) = \ractie^M_m(\xi)$.
\item The map def\/ined pointwise as the dual to $\ractie^M_m$ is denoted by $(\ractie^M)^*: T^*M \to \lag^*$.
\item The lifted action $\Psi^{TM}$ on $TM$ of $\Psi^M$ is given by $\Psi^{TM}: G\times TM \to TM; (g,v_m) \mapsto T\Psi^M_g(v_m)$.
\item The lifted action $\Psi^{T^*M}$ on $T^*M$ is given by $\Psi^{T^*M}: G\times T^*M \to T^*M; (g,\alpha_m) \mapsto T^*\Psi^M_{g^{-1}}(\alpha_m)$.
\end{enumerate}
\end{definition}
With these notations, $\xi_{TM}$ denotes a fundamental vector f\/ield on $TM$ determined by the lifted action. It follows that $\xi_{TM}$ is the complete lift of $\xi_M$.

Every right action $\Psi^M$ gives rise to a left action $\Phi^M$: $\Phi^M_{g}(m) = \Psi^M_{g^{-1}}(m)$. We only consider right actions. This is not a true restriction since one may reformulate the main results for left actions if needed.

The orbit space $M/G$ of a free and proper action is a manifold and $\pi:M\to M/G$ carries the structure of a principal $G$-bundle. Throughout the paper $[m]_G$ denotes a point in the orbit space $M/G$, i.e.\ it is the orbit through $m\in M$.  Every tangent vector in the kernel of $T\pi$ is of the form $\xi_M(m)$ for some $\xi\in\lag$. These vectors are called vertical and form a distribution which we call the vertical distribution $V\pi=\ker T\pi$.

\begin{definition} A principal connection on a principal $G$-bundle is a $\lag$-valued 1-form $\A$ satisfying two conditions:
\begin{enumerate}\itemsep=0pt
\item[1)] it is equivariant, i.e.\ $\Psi^*_g\A = {\rm Ad}_{g^{-1}} \cdot \A$ for any $g\in G$, and
\item[2)] for $\xi\in\lag$ arbitrary, $\A ( \xi_M)=\xi$.\end{enumerate} \end{definition}
The kernel of $\A$ determines a right invariant distribution on $M$ which is a complement of the vertical distribution. It is therefore called the {\em horizontal} distribution of $\A$ and is typically denoted by $H^{\A}\subset TM$. On the other hand, any right invariant distribution $H$ satisfying $H\oplus V\pi = TQ$ determines a principal connection, see~\cite{koba}.

We denote by $\langle f ,\A\rangle$, where $f$ is a $\lag^*$-valued function on $M$,  the 1-form on $M$ pointwise def\/ined by \[v_m\mapsto \langle f(m),\A(m)(v_m)\rangle\in \R.\] The cotangent vector $\langle f ,\A\rangle(m)\in T_m^*M$ is often denoted by $\langle f(m),\A(m)\rangle$.  In particular, if $f=\mu\in\lag^*$ is constant, then $\langle \mu,\A\rangle$ is a 1-form on $M$.

{\bf Routh reduction.}

\begin{definition} Let $(Q,L)$ denote a Lagrangian system and assume that the conf\/iguration space is equipped with an action $\Psi^Q$. The Lagrangian system $(Q,L)$ is $G$-invariant if $L$ is invariant under the lifted action $\Psi^{TQ}$, i.e.\ $L(\Psi^{TQ}_g(v_q))=L(v_q)$ for arbitrary $v_q\in TQ$ and $g\in G$.
\end{definition}

If a mechanical Lagrangian system is $G$-invariant then the kinetic energy metric $\ll \cdot, \cdot \gg_Q$ and the potential energy $V$ are both invariant under the pull-back of $\Psi^Q_g$, for arbitrary $g$.

\begin{definition}\mbox{}
\begin{enumerate}\itemsep=0pt
\item For a Lagrangian system $(Q,L)$, the Legendre transform $\F L:TQ\to T^*Q$ is the f\/ibre derivative of $L$, i.e.\ for arbitrary $v_q,w_q\in TQ$
\[\langle \F L(v_q), w_q\rangle  = \left.\frac{d}{du}\right|_{u=0} L(v_q+u w_q).\]
The Lagrangian is hyperregular if $\F L$ is a dif\/feomorphism.
\item For a $G$-invariant Lagrangian system $(Q,L)$, the momentum map $J_L : TQ\to \lag^*$ is the map $(\psi^Q)^*\circ \F L$, i.e.\ for arbitrary  $v_q\in TQ$ and $\xi \in \lag$
\[\langle J_L(v_q),\xi\rangle = \langle \F L(v_q) , \xi_{Q}(q)\rangle.\]
\item Given a $G$-invariant mechanical Lagrangian system $(Q,L)$, and a point $q$ in $Q$. The inertia tensor $\mathbb{I}_q$ is a metric on $\lag$ def\/ined by $\mathbb{I}_q(\xi,\eta)= \ll \xi_Q(q),\eta_Q(q)\gg_Q$.
\end{enumerate}
\end{definition}
The momentum map associated to a $G$-invariant Lagrangian system satisf\/ies the following two important properties:
\begin{enumerate}\itemsep=0pt \item It is conserved along the solutions of the Euler--Lagrange equations, i.e.\ if $q(t)$ is a solution to the Euler--Lagrange equations, then $\frac{d}{dt} \left(J_L(\dot q(t))\right)=0$.
\item It is equivariant w.r.t.\ the action $\Psi^{TQ}$ and the coadjoint action ${\rm Ad}^*$ on $\lag^*$, i.e.\ \[J_L(\Psi^{TQ}_g(v_q)) = {\rm Ad}^*_g(J_L(v_q))\] for arbitrary $g\in G$ and $v_q\in TQ$.
\end{enumerate}

\begin{definition} Given a $G$-invariant Lagrangian system $(Q,L)$ and an arbitrary vector $v_q\in TQ$.{\samepage
\begin{enumerate}\itemsep=0pt
\item The map $J_L|_{v_q}: \lag \to \lag^*$ is the map $\xi \mapsto J_L(v_q+\xi_Q(q))$.
\item The Lagrangian system $(Q,L)$ is  $G$-regular if $J_L|_{v_q}$ is a dif\/feomorphism for every $v_q\in TQ$.
\end{enumerate}}
\end{definition}

Every $G$-invariant mechanical Lagrangian system is $G$-regular. To show this, remark that for a mechanical Lagrangian system \[\langle J_L|_{v_q}(\xi)  , \eta \rangle =\ll v_q,\eta_Q(q)\gg_Q+ \ll \xi_Q(q), \eta_Q(q)\gg_Q.\] Hence $J_L|_{v_q}(\xi) = J_L|_{v_q}(0) + \mathbb{I}_q(\xi)$ is an af\/f\/ine map, whose linear part is determined by the inertia metric on $\lag$.
\begin{proposition}
Consider a $G$-invariant and $G$-regular Lagrangian system $(Q,L)$ and fix a~re\-gu\-lar value $\mu\in\lag^*$ of the momentum map $J_L$. Let $G_\mu$ denote the isotropy subgroup of $\mu$ w.r.t.\ the coadjoint action of $G$ on $\lag^*$, i.e.\ $g\in G_\mu$ iff ${\rm Ad}^*_g\mu = \mu$.
\begin{enumerate}\itemsep=0pt
\item[$1.$] The submanifold $i_\mu:J^{-1}_L(\mu)\to TQ$ is $G_\mu$-invariant and the restricted action of $G_\mu$ on~$J^{-1}_L(\mu)$ is free and proper.
\item[$2.$] The quotient manifold $J^{-1}_L(\mu)/G_\mu$ is diffeomorphic to the fibred product $T(Q/G) \times_{Q/G} Q/G_\mu$ over~$Q/G$.
\end{enumerate}
\end{proposition}
The f\/irst statement is a direct consequence of the equivariance of $J_L$. We postpone a proof of the second statement (Proposition~\ref{prop:gregular}, page~\pageref{prop:gregular}).

We are now ready to describe a preliminary version of Routh reduction where the symmetry group $G$ is Abelian and the bundle $Q\to Q/G$ is trivial, so that we can choose a connection $\A$ with vanishing curvature.  Since $G$ is Abelian, we have that the isotropy subgroup $G_\nu$ for every $\nu \in \lag^*$ is the entire group $G$, and in particular $T(Q/G)\times_{Q/G} Q/G_\nu$ is nothing but $T(Q/G)$.
\begin{theorem}[Routh reduction -- Abelian version] Let $G$ be Abelian and let $\A$ be a connection on $Q$ with vanishing curvature. The Routh reduction procedure of a $G$-invariant and $G$-regular Lagrangian system $(Q,L)$ consists of the following steps.
\begin{enumerate}\itemsep=0pt
\item[$1.$] Fix a regular value $\mu$ of the momentum map $J_L$ and consider the submanifold $
J^{-1}_L(\mu)$.
\item[$2.$] Compute the restriction of the $G$-invariant function $L-\langle \mu , \A \rangle$ to the level set $J^{-1}_L(\mu)$. Let $\tilde L$ denote its projection to the quotient manifold $T(Q/G)\cong J^{-1}_L(\mu)/G$.
\end{enumerate}
Every solution of the Euler--Lagrange equations of the Lagrangian system $(Q/G,\tilde L)$ is the projection of a solution of the Euler--Lagrange equations of the original system~$(Q,L)$ with momentum~$\mu$. Conversely, every solution of the Euler--Lagrange equations of the Lagrangian system~$(Q,L)$ with momentum~$\mu$ projects to a solution of the Euler--Lagrange equations of the system~$(Q/G,\tilde L)$.
\end{theorem}
One may also f\/ind a description of global Abelian Routh reduction in~\cite{arnold}.

{\bf Example: the spring pendulum.} The system consists of a point particle with mass~$m$ moving in a horizontal plane, and attached to the origin by means of a spring with spring constant~$k$.  We choose polar coordinates $(r,\theta)$ for this system, so that the mechanical Lagrangian is given by $L = \frac12 m (\dot r^2+ r^2\dot\theta^2)  - \frac12 k r^2$.  This system is clearly invariant under translations in the $\theta$-direction, given by  $\Psi_{a}(r,\theta) = (r,\theta+a)$. The momentum map for this action is $J_L=mr^2\dot\theta$ and the Euler--Lagrange equations are
\begin{gather*} m\ddot r - mr\dot\theta^2+kr =0,\\
\frac{d}{dt} (mr^2\dot\theta)= 0.
\end{gather*}
Let us f\/ix a regular value $0\neq\mu=mr^2\dot\theta$ for the momentum map and let $\A = d\theta$ be the standard connection with vanishing curvature. The Routhian is obtained from \[\tilde L(r,\dot r)=(L-\mu\dot\theta)|_{mr^2\dot\theta=\mu} = \frac12 m\dot r^2 -\frac12 k r^2 - \frac12\frac{\mu^2}{mr^2}.\]
The Euler--Lagrange equation for $\tilde L$ is $m\ddot r =- kr + \frac{\mu^2}{mr^3}$ and solutions of this equations are in correspondence to solutions of the Euler--Lagrange equations for $L$ with momentum $\mu$.

Routh reduction can be extended to more general contexts, including non-Abelian group actions~\cite{marsdenrouth},  Lagrangians of non-mechanical type~\cite{adamec,mestcram},   Lagrangians invariant up to a total time derivative~\cite{BC} and  Lagrangians that are not $G$-regular~\cite{BM}.  Additional complications arise in these cases: for instance, if the bundle $Q\to Q/G$ is not trivial, the reduced system is typically subjected to an additional force term associated to the curvature of the chosen connection. On the other hand, if the symmetry group is not Abelian, the quotient space $J^{-1}_L(\mu)/G_\mu$ no longer has the structure of a tangent bundle; instead it is dif\/feomorphic to $T(Q/G)\times_{Q/G} Q/G_\mu$. The interpretation of the reduced system as a Lagrangian system then requires additional def\/initions, which are postponed to the following section.

Our main interest in this paper is Routh reduction by stages, where the assumption is that the Lagrangian system obtained after applying a f\/irst Routh reduction carries additional symmetry, so that we can reapply Routh reduction. The system obtained after one Routh reduction is formulated on a f\/ibred product of the type $T(Q/G)\times_{Q/G} Q/G_\mu$. This is the f\/ibred product of a~bundle $Q/G_\mu \to Q/G$ with the tangent bundle to the base space. This observation is our main motivation to extend Routh reduction to Lagrangian systems def\/ined on such f\/ibred products:  in order to develop Routh reduction by stages, we have to be able to reduce Lagrangian systems obtained after a f\/irst reduction.

Routh reduction is closely related to symplectic reduction on the associated cotangent bundles (see~\cite{BC,marsdenrouth}). In the next section we will introduce the concept of a ``magnetic Lagrangian system'' and we will emphasize its symplectic formulation. The concept of magnetic Lagrangian systems is the analogue of Hamiltonian systems one encounters in magnetic cotangent bundle reduction~\cite{MarsdenHamRed}.

\section{Magnetic Lagrangian systems}\label{section3}

A magnetic Lagrangian system is a Lagrangian system with conf\/iguration space the total space of a bundle $\e:P\to Q$ and where the Lagrangian is independent of the velocities tangent to the f\/ibres of $\e$. Additionally the system is subjected to a force term that is of magnetic type. It might help to keep in mind that in the case of a Routh reduced Lagrangian system, $P$ corresponds to~$Q/G_\mu$ and the f\/ibration $\e$ is given by the projection $Q/G_\mu \to Q/G$.

\begin{definition} A magnetic Lagrangian system is a triple $(\e:P\to Q, L, \B)$ where $\e:P\to Q$ is a f\/ibre bundle, $L$ is a smooth function on the f\/ibred product $TQ\times_Q P$ and $\B$ is a closed 2-form on $P$. We say that $P$ is the conf\/iguration manifold of the system and that $L$ is the Lagrangian. \end{definition}

A coordinate chart $(q^i,p^a)$, $i=1,\ldots,n=\dim Q$, $a=1,\ldots,k=\dim P-\dim Q$,  adapted to the f\/ibration $\e: P\to Q$ determines a coordinate chart $(q^i,v^i,p^a)$ on $TQ\times_Q P$, and the Lagrangian~$L$ is then a function depending on $(q^i,v^i,p^a)$. By def\/inition, $L$ is independent of the velocities in the f\/ibre coordinates $p^a$ and therefore it determines a singular Lagrangian when interpreted as a function on $TP$. Locally, the Euler--Lagrange equations for this singular Lagrangian are
\begin{gather*}
\frac{d}{dt}\left(\fpd{L}{v^i}\right) -\fpd{L}{q^i}=\B_{ij} \dot q^j + \B_{ia} \dot p^a,\qquad  i=1,\ldots,n,\\
  -\fpd{L}{p^a}=-\B_{ia} \dot q^i+ \B_{ab}\dot p^b,\qquad a=1,\ldots,k.
\end{gather*}
Here we used the following coordinate expression of the 2-form $\B$ is $\frac12 \B_{ij} dq^i \wedge dq^j + \B_{ia}dq^i\wedge dp^a + \frac12 \B_{ab} dp^a\wedge dp^b$.
These Euler--Lagrange equations have a geometric interpretation (see Proposition~\ref{def:el}). First we introduce additional notations.
\begin{definition}
Assume a magnetic Lagrangian system $(\e:P\to Q, L,\B)$ is given.
\begin{enumerate}\itemsep=0pt
\item   $T_PQ$ denotes the f\/ibred product $TQ\times_Q P$ and $(v_q,p)$, where $v_q\in TQ$ and $p\in P$ such that $\e(p)=q$, is a point in $T_PQ$. Similarly, $T_P^*Q$ denotes the f\/ibred product $T^*Q\times_Q P$ and $(\alpha_q,p)$, with $\e(p)=q$, is an arbitrary element in $T^*_PQ$.
\item $V\e$ denotes the distribution on $P$ of tangent vectors vertical to $\e$.
\item $\hat \e : TP \to T_PQ$ is the projection f\/ibred over $P$ that maps $v_p\in TP$ onto $(T\e(v_p),p)\in T_PQ$.
\item $\rho_1:T_PQ \to TQ$ is the projection that maps $(v_q,p )\in T_PQ$ onto $v_q\in TQ$.
\item $\rho_2: T_PQ \to P$ is the projection that maps $(v_q,p)\in T_PQ$ onto $p\in P$.
\item $\e_1:T_P^*Q \to T^*Q$ is the projection that maps $(\alpha_q,p )\in T_P^*Q$ onto $\alpha_q\in T^*Q$.
\item $\e_2: T_P^*Q \to P$ is the projection that maps $(\alpha_q,p)\in T_P^*Q$ onto $p\in P$.
\item The Legendre transform $\F L:T_PQ \to T_P^*Q$ maps $(v_q,p)\in T_PQ$ to $(\alpha_q,p)\in T^*_PQ$ where $\alpha_q\in T^*_qQ$ is determined from \[\langle \alpha_q, w_q\rangle = \left.\frac{d}{du}\right|_{u=0} L(v_q+uw_q,p),\] for arbitrary $w_q\in T_qQ$.
\item The energy $E_L$ is a function on $T_PQ$ def\/ined by $E_L(v_q,p) =\langle \F L (v_q,p), (v_q,p)\rangle - L(v_q,p)$. (Here the contraction of an element $(\alpha_q,p)\in T^*_PQ$ with $(v_q,p)\in T_PQ$ is def\/ined naturally as $\langle (\alpha_q,p),(v_q,p) \rangle =\langle \alpha_q,v_q\rangle$.)
\item By means of the Legendre transform we can pull-back the presymplectic 2-form $\e^*_1\omega_{Q}+\e^*_2\B$ on $T^*_PQ$ to a presymplectic 2-form $\F L^*(\e^*_1\omega_{Q} +\e^*_2\B)$ on $T_PQ$. The latter is denoted by $\Omega^{L,\B}$ (Here $\omega_{Q}=d\theta_{Q}$, with $\theta_{Q}$ the Poincar\'e--Cartan 1-form on $T^*Q$ and a presymplectic 2-form is understood to be a closed 2-form, not necessarily of constant rank.)
\end{enumerate}
\end{definition}

\begin{proposition}\label{def:el} Given a curve $p(t)$ in $P$, and let $\gamma(t)$ denote the curve in $T_PQ$ equal to $(\dot q(t), p(t))\in T_PQ$ with $q(t)=\e(p(t))$. The curve $p(t)$ in $P$ is a solution to the Euler--Lagrange equations for the magnetic Lagrangian system $(\e:P\to Q,L,\B)$ iff  $\gamma(t)$ is a solution to the presymplectic equation
\[i_{\dot \gamma(t)} \Omega^{L,\B} (\gamma(t)) = -dE_L(\gamma(t)).\]
\end{proposition}
Locally, the presymplectic equation coincides with the previously mentioned Euler--Lagrange equations, since
\begin{gather*}
\Omega^{L,\B}  = d\left(\fpd{L}{v^i} \right)\wedge d q^i + \frac12 \B_{ij} dq^i \wedge dq^j + \B_{ia}dq^i\wedge dp^a + \frac12 \B_{ab} dp^a\wedge dp^b,\\
 dE_L   =v^i d\left(\fpd{L}{v^i} \right) + \fpd{L}{v^i} dv^i - dL=v^i d\left(\fpd{L}{v^i} \right) - \fpd{L}{q^i} dq^i - \fpd{L}{p^a} dp^a.
\end{gather*}

\begin{definition}A magnetic Lagrangian system $(\e:P\to Q,L,\B)$
\begin{enumerate}\itemsep=0pt \item[1)] is hyperregular if $\F L$ is  a dif\/feomorphism and if the restriction of $\B$ to $V\e$ is nondegenerate,
 \item[2)] is of mechanical type if $L(v_q,p)= \frac12 \ll (v_q,p) ,(v_q,p)\gg_{\rho_1}- V(p)$ with $\ll \cdot , \cdot \gg_{\rho_1}$ is a metric on the vector bundle $\rho_1:T_PQ \to P$ and $V$ is a function on $P$. \end{enumerate}
\end{definition}

Note that if $\mathcal{B}$ is nondegenerate, then the typical f\/ibre of $P$ necessarily has to be even-dimensional.
In a local coordinate chart the nondegeneracy condition on $\B$ is expressed by $\det \B_{ab}\neq0$.
\begin{proposition} If a magnetic Lagrangian system $(\e:P\to Q,L,\B)$ is hyperregular, the $2$-form $\Omega^{L,\B} = \F L^*(\e^*_1\omega_{Q} +\e^*_2\B)$ determines a symplectic structure on~$T_PQ$.
\end{proposition}
\begin{proof}
Assume that the magnetic Lagrangian system is hyperregular. The nondegeneracy of~$\Omega^{L,\B}$ is easily checked if we work in a coordinate chart adapted to the f\/ibration:
\[
\Omega^{L,\B} = d\left(\fpd{L}{v^i} \right)\wedge d q^i + \frac12 \B_{ij} dq^i \wedge dq^j + \B_{ia}dq^i\wedge dp^a + \frac12 \B_{ab} dp^a\wedge dp^b.
\]
Since $\F L$ is a dif\/feomorphism, the 1-forms $d\left(\fpd{L}{v^i} \right)$, $dq^i$ and $dp^a$ provide pointwise a basis on~$T^*(T_PQ)$. One can now use standard arguments to prove the nondegeneracy.
\end{proof}

We conclude that a hyperregular magnetic Lagrangian system has a symplectic structure although the Lagrangian itself is singular when interpreted as a function on $TP$. The energy~$E_L$ is the Hamiltonian. This is an important observation.

\begin{remark} Throughout the paper we only consider magnetic Lagrangian systems that are hyperregular. This is not a true restriction. The results remain valid for general magnetic Lagrangian systems: instead of relating Routh reduction to symplectic reduction, it is possible to relate it to presymplectic reduction~\cite{presympred}. \end{remark}

\begin{remark} We conclude this section with a remark on equivalent magnetic Lagrangian systems. Roughly speaking, two Lagrangian systems are equivalent if the resulting dynamics coincide, more specif\/ically if they produce the same Euler--Lagrange equations. It is well known that for a Lagrangian system the Euler--Lagrange equations do not change when the Lagrangian is augmented with a total time derivative of a function on the conf\/iguration space. For magnetic Lagrangian systems this gauge freedom can be extended.

Consider a 1-form $\alpha$ along $\e$, i.e.\ a section of $\e_2:T^*_PQ\to P$, or in other words a linear function on $T_PQ$ given by $(v_q,p)\mapsto \langle \alpha(p),(v_q,p)\rangle$. When $\alpha:P\to T^*_PQ$ is composed with $T^*\e\circ\e_1:T^*_PQ\to T^*P$, it determines a 1-form on $P$ which, with a slight abuse of notation, we denote by $\e^*\alpha$. Now consider the function $L'(v_q,p)= L(v_q,p)- \langle \alpha(p),(v_q,p)\rangle$ and the gyroscopic 2-form $\B'= \B +d\e^*\alpha$.  Together they def\/ine a new magnetic Lagrangian system $(\e:P\to Q, L',\B ')$ which is equivalent to $(\e:P\to Q,L,\B)$.
\end{remark}

\begin{lemma}
 The magnetic Lagrangian systems $(\e, L',\B ')$ and $(\e,L,\B)$ are equivalent.
\end{lemma}

This is a reformulation of a well-known result in classical mechanics, saying that an exact gyroscopic force can be taken into account by means of a velocity dependent potential~\cite{pars}.
\begin{proof}
We show that any solutions to the Euler--Lagrange equations of $(\e, L',\B ')$ is a solution to the Euler--Lagrange equations of $(\e,L,\B)$. We work in a local coordinate neighborhood $(q^i,p^a)$ as before. We f\/ix a coordinate expression for $\alpha=\alpha_i(q^j,p^a)dq^i$. The Euler--Lagrange equations for $(\e,L',\B')$ equal, with $\dot q^i= v^i$ and $\dot p^a= v^a$
\begin{gather*}
\frac{d}{dt}\left(\fpd{L'}{v^i}\right) -\fpd{L'}{q^i}=\B'_{ij}v^j + \B'_{ia} v^a,\qquad i=1,\ldots,n,\\
  -\fpd{L'}{p^a}=-\B'_{ia}v^i + \B'_{ab} v^b,\qquad a=1,\ldots,k.
\end{gather*}
It now remains to substitute the def\/inition of $L'$ and $\B'$, i.e.\ $L'= L - \alpha_i v^i$ and
\begin{gather*}
\B'_{ij} = \B_{ij} + \left(\fpd{\alpha_j}{q^i}-\fpd{\alpha_i}{q^j}\right), \qquad \B'_{ia} = \B_{ia} - \fpd{\alpha_i}{p^a}, \qquad \B'_{ab} = \B_{ab}. \tag*{\qed}
\end{gather*}
\renewcommand{\qed}{}\end{proof}

\section{Magnetic Lagrangian systems with symmetry}\label{section4}

\subsection{Symplectic reduction: a brief introduction}\label{section4.1}

An overall reference for this section is e.g.~\cite{MarsdenHamRed}. Let $(M,\omega)$ be a symplectic manifold on which $G$ acts on the right, $\Psi^M:M\times G\to M$. Given a function $f: M\to \lag^*$, then $f_\xi$ for $\xi\in\lag$ denotes the real valued function on $M$ def\/ined by $f_\xi(m)= \langle f(m),\xi\rangle$.
\begin{definition}\label{def:cocycle}\mbox{}
\begin{enumerate}\itemsep=0pt
\item The action $\Psi^M$  is said to be {\em canonical} if $(\Psi^M_g)^*\omega=\omega$ for all $g\in G$.
\item A map $J:M\to \lag^*$ is a momentum map if $i_{\xi_M}\omega=-dJ_\xi$, for $\xi \in\lag$ arbitrary.
\item If $M$ is connected, the {\em non-equivariance $1$-cocycle} $\sigma$ of the momentum map $J$ equals
\[
\sigma: \ G\to \lag^*: \ g\mapsto J\big(mg^{-1}\big)-{\rm Ad}^*_{g^{-1}}(J(m)),
\]
where $m$ is arbitrary in $M$.
\end{enumerate}\end{definition}
The def\/inition of $\sigma$ is independent of $m$ (see~\cite{Marsden}). Recall that a 1-cocycle with values in $\lag^*$ statisf\/ies, for $g,h\in G$ arbitrary,
\[
\sigma(gh)=\sigma(g)+ {\rm Ad}^*_{g^{-1}}\sigma(h).
\]

\begin{definition} Let $\sigma$ be a 1-cocycle with values in $\lag^*$. The af\/f\/ine action of $G$ on $\lag^*$ with 1-cocycle $\sigma$ is given by $(g,\mu)\mapsto {\rm Ad}^*_{g}\mu +\sigma(g^{-1})$ for arbitrary $\mu\in\lag^*$ and $g\in G$.
\end{definition}
The momentum map $J$ is equivariant with respect to the af\/f\/ine action with 1-cocyle from Definition~\ref{def:cocycle}: $J(mg)= {\rm Ad}_g^* J(m)+ \sigma(g^{-1})$. For a f\/ixed element $\mu\in\lag^*$, the Lie group $G_\mu<G$ denotes the isotropy subgroup of $\mu$ w.r.t.\ the af\/f\/ine action, i.e.\ $g\in G_\mu$ if $\mu = {\rm Ad}_{g}^*\mu + \sigma(g^{-1})$ or equivalently if \[\mu - J(mg) = {\rm Ad}_g^*(\mu - J(m)).\]

\begin{theorem}[symplectic reduction SR]\label{thm:mw}
Let $(M,\omega)$ be a symplectic manifold, with $G$ acting canonically on $M$. Let $J$ be a momentum map for this action with non-equivariance cocycle $\sigma$. Assume that $\mu$ is a regular value of $J$, and denote by $G_\mu$ the isotropy group of $\mu$ under the affine action of $G$ on $\lag^*$. Then the pair $(M_\mu,\omega_\mu)$ is a symplectic manifold, with $M_\mu = J^{-1}(\mu)/G_\mu$ and with $\omega_\mu$ a $2$-form on $M_\mu$ uniquely determined from $i^*_\mu \omega=\pi_\mu^*\omega_\mu$, with  $i_\mu:J^{-1}(\mu)\to M$ and $\pi_\mu: J^{-1}(\mu)\to M_\mu=J^{-1}(\mu)/G_\mu$.

Any Hamiltonian $h$ on $M$ which is invariant under the action of $G$ induces a function $h_\mu$ on $M_\mu$ satisfying $\pi_\mu^*h_\mu = i^*_\mu h$. The Hamiltonian vector field~$X_h$ is tangent to $J^{-1}(\mu)$ and the corresponding vector field on $J^{-1}(\mu)$ is $\pi_\mu$-related to the Hamiltonian vector field $X_{h_\mu}$ on $M_\mu$.
\end{theorem}

\subsection{Invariant magnetic Lagrangian systems and momentum maps}\label{section4.2}
In order to def\/ine invariant magnetic Lagrangian systems, we start from a free and proper action of a Lie group $G$ on both $P$ and $Q$ such that they commute with $\e$: for arbitrary $g\in G$,
\begin{gather}\label{compatiblea}\e\circ\Psi^P_g=\Psi^Q_g\circ\e.\end{gather}
In other words, $G$ acts on $\epsilon: P \to Q$ by bundle automorphisms.
The projections of the principal bundles are denoted by $\pi^Q:Q\to Q/G$ and $\pi^P:P\to P/G$.
These actions induce lifted right actions on $T_PQ$ and $T_P^*Q$:
\begin{enumerate}\itemsep=0pt
\item[1)] $\Psi^{T_PQ}_g(v_q,p):= (T\Psi^Q_g(v_q),\Psi^P_g(p))$ for $(v_q,p)\in T_PQ$ and $g\in G$,
\item[2)] $\Psi^{T^*_PQ}_g(\alpha_q,p):= (T^*\Psi^Q_{g^{-1}}(\alpha_q),\Psi^P_g(p))$ for $(\alpha_q,p)\in T^*_PQ$ and $g\in G$.
\end{enumerate}

\begin{definition}  A magnetic Lagrangian system $(\e:P\to Q,L,\B)$ is $G$-invariant if
\begin{enumerate}\itemsep=0pt
\item[1)] $\B$ is invariant under the action of $\Psi^P$: $(\Psi^P_g)^*\B = \B$ for all $g\in G$,
\item[2)] the Lagrangian $L$ is invariant under the lifted action $\Psi^{T_PQ}$ on $T_PQ$, i.e.\ $L\circ \Psi^{T_PQ}_g = L$ for all $g\in G$.
    \end{enumerate}
\end{definition}

Let  $(\e:P\to Q, L,\B)$ be a $G$-invariant hyperregular magnetic Lagrangian system and consider the symplectic 2-form $\e^*_1\omega_Q +\e^*_2 \B$ on $T^*_PQ$.
\begin{proposition}\label{prop:equivleg} The action $\Psi^{T^*_PQ}$ is canonical and the Legendre transform $\F L: T_PQ\to T_P^*Q$ is equivariant, i.e.\  $\F L \circ \Psi^{T_PQ}_g=\Psi_g^{T^*_PQ}\circ \F L$. As a consequence the action $\Psi^{T_PQ}$ is canonical w.r.t.~$\Omega^{L,\B}$.
\end{proposition}
\begin{proof}
Note that $\e_1 \circ \Psi^{T^*_PQ}_g = T^*\Psi^Q_{g^{-1}}\circ \e_1$ and $\e_2 \circ \Psi^{T^*_PQ}_g = \Psi^P_g\circ \e_2$. Since the action $\Psi^{T^*Q}_g=T^*\Psi^Q_{g^{-1}}$ on $T^*Q$ is canonical w.r.t.\ $\omega_Q$ and since $\B$ is a $G$-invariant 2-form on $P$, we conclude that $\Psi^{T^*_PQ}$ is canonical. To show the equivariance of $\F L$ we use its def\/inition: let $g\in G$, $v_q,w_q\in TQ$ be arbitrary, then
\begin{gather*}
\langle \F L\left(\Psi_g^{T_PQ}(v_q,p)\right) , \Psi_g^{T_PQ} (w_q,p) \rangle  = \left.\frac{d}{du}\right|_{u=0} L( \Psi_g^{T_PQ}(v_q+uw_q))\\
\hphantom{\langle \F L\left(\Psi_g^{T_PQ}(v_q,p)\right) , \Psi_g^{T_PQ} (w_q,p) \rangle}{}
=\left.\frac{d}{du}\right|_{u=0} L(v_q+uw_q) = \langle \F L(v_q,p), (w_q,p)\rangle. \tag*{\qed}
\end{gather*}\renewcommand{\qed}{}
\end{proof}

In order to def\/ine a momentum map, we introduce a concept that is similar to the so-called  $\B\lag$-potential in~\cite{MarsdenHamRed}.

\begin{definition} \label{def:bgpotential}
A $\lag^*$-valued function $\delta$ on $P$ is a $\B\lag$-potential for the magnetic 2-form $\B$ if $i_{\xi_{P}}\B = d\delta_{\xi}$  for any $\xi \in \lag$.
\end{definition}

Recall that $\delta_\xi$ is def\/ined pointwise as $\delta_\xi(p) = \langle \delta(p),\xi\rangle$. If $\B$ is $G$-invariant and admits a~$\B\lag$-potential $\delta$, then it follows that $d((\Psi^P_g)^* \delta_\xi - \delta_{{\rm Ad}_g\xi})=0$. If $P$ is connected, this induces a~$\lag^*$-valued 1-cocycle on $G$:
\[
\sigma_\delta: \ G\to \lag^{*}; \qquad g \mapsto (\Psi^{P}_{g^{-1}})^{*}\delta(p) - {\rm Ad}^{*}_{g^{-1}}\cdot \delta(p),
\]
with $p$ arbitrary in $P$. The inf\/initesimal version of this 1-cocycle is denoted by $\Sigma_\delta(\xi,\eta)=-\langle T_e\sigma_\delta(\xi),\eta\rangle$. It is a 2-cocycle on the Lie algebra and satisf\/ies  $\Sigma_\delta(\xi,\eta)= -\B(\xi_{P},\eta_{P})-\delta_{[\xi,\eta]} = \xi_P(\delta_\eta)-\delta_{[\xi,\eta]}$. For the following proposition, recall from Def\/inition~\ref{def:acties} that $(\ractie^Q)^*: T^*Q \to \lag^*$ is the dual of the inf\/initesimal action of $G$ on $Q$.

\begin{proposition}\label{prop:equiv} The map $J_\delta=(\ractie^Q)^*\circ\e_{1}-\delta\circ\e_{2}$ is a momentum map for the symplectic manifold $(T_P^*Q,\e^*_1\omega_Q +\e^*_2 \B)$ with non-equivariance $1$-cocycle $-\sigma_\delta$. Due to the equivariance of~$\F L$, the map $J_{L,\delta}= J_\delta\circ \F L$ is a momentum map for the symplectic manifold $(T_PQ,\Omega^{L,\B})$ with non-equivariance $1$-cocycle $-\sigma_\delta$. \end{proposition}
\begin{proof}
The map $(\ractie^Q)^*: T^*Q\to \lag^*$ is an equivariant momentum map for the symplectic manifold $(T^*Q,\omega_Q)$. It is straightforward that the combined map $J_\delta=(\ractie^Q)^*\circ\e_{1}-\delta\circ\e_{2}$ is a momentum map for the lifted action on $T^{*}_{P}Q$ w.r.t.\ the symplectic form $\e_{1}^{*}\omega_{Q} + \e_{2}^{*}\B$ with non-equivariance cocycle $-\sigma_\delta$. The rest of the statement follows by construction.
\end{proof}

We conclude that the symplectic structures $(T_PQ,\Omega^{L,\B})$ and $(T_P^*Q,\e^*_1\omega_Q +\e^*_2 \B)$ associated to a $G$-invariant and hyperregular magnetic Lagrangian system with a $\B\lag$-potential $\delta$ admit a~momentum map with cocycle $-\sigma_\delta$ and are amenable to symplectic reduction.

\section{Routh reduction for magnetic Lagrangian systems}\label{section5}
\subsection{Magnetic cotangent bundle reduction}\label{section5.1}

In this section we study the reduction of the symplectic manifold $(T_P^*Q, \e_1^*\omega_Q +\e^*_2\B)$ determined from the magnetic 2-form $\B$ of a $G$-invariant hyperregular magnetic Lagrangian system with $\B\lag$-potential $\delta$. From Proposition~\ref{prop:equiv}, it follows that the function $J_\delta$ is a momentum map with non-equivariance cocycle $-\sigma_\delta$.  As usual, $G_\mu$ denotes the isotropy subgroup of $\mu\in\lag^*$ for the af\/f\/ine action.

\begin{proposition} Fix a connection $\A^Q$ on the bundle $\pi^Q: Q\to Q/G$ and a regular value $\mu$ of the momentum map $J_\delta$.
\begin{enumerate}\itemsep=0pt
\item[$1.$] $g\in G_\mu$ iff $\mu+\delta(pg)= {\rm Ad}_g^*(\mu+\delta(p))$.
\item[$2.$] $\A^P=\e^*\A^Q$ is a principal connection on $\pi^P:P\to P/G$.
\item[$3.$] The $1$-form $\langle \mu+\delta,\A^P\rangle$ on $P$ is $G_\mu$-invariant.
\item[$4.$]  The $2$-form $\B + d(\langle \mu+\delta,\A^P\rangle)$ is $G_\mu$-invariant and reducible to a $2$-form on $P/G_\mu$.
\item[$5.$] The quotient manifold $J_\delta^{-1}(\mu)/G_\mu$ is diffeomorphic to $T^*_{P/G_\mu}(Q/G)$.
\end{enumerate}
\end{proposition}

\begin{proof} Recall the pointwise def\/inition of $\langle \mu +\delta,\A^P\rangle$: given $v_p\in TP$, then $\langle \mu +\delta,\A^P\rangle(p)(v_p) := \langle  \mu +\delta(p),\A^P(p)(v_p)\rangle$. Below, $\langle \mu +\delta(p),\A^P(p)\rangle\in T^*_pP$ denotes the cotangent vector $\langle \mu +\delta,\A^P\rangle(p)$. We continue with the proof.

1. This is straightforward from the def\/inition of $\sigma_\delta$.

 2. The $\lag^*$-valued 1-form $\A^P$ determines a principal connection if it satisf\/ies $\A^P(\xi_P)=\xi$ for all $\xi\in\lag$, and if $(\Psi^P_g)^*\A^P={\rm Ad}_{g^{-1}}\cdot \A^P$ for all $g\in G$:
\begin{gather*}
  \A^P(\xi_P)(p) = \A^Q(\e(p))(T\e(\xi_P(p)))= \A^Q(\e(p))(\xi_Q(\e(p))) = \xi \qquad \mbox{and}\\
 (\Psi^P_g)^*\A^P= (\Psi^P_g)^*\e^*\A^Q = \e^* (\Psi^Q_g)^*\A^Q = {\rm Ad}_{g^{-1}}\cdot \e^*\A^Q ={\rm Ad}_{g^{-1}}\cdot\A^P.
\end{gather*}

3. The pull-back 1-form $(\Psi^P_g)^*\langle \mu +\delta,\A^P\rangle$ equals
\[
 \langle \mu +\delta(pg),((\Psi^P_g)^*\A^P)(p)\rangle=\langle {\rm Ad}_g^*( \mu +\delta(p)),{\rm Ad}_{g^{-1}}\cdot\A^P(p)\rangle.
\]

 4. That $\B + d(\langle \mu +\delta,\A^P\rangle)$ is $G_\mu$-invariant is a straightforward consequence of~3. It is projectable to $P/G_\mu$ since the contraction with any fundamental vector f\/ield $\xi_P$ with $\xi\in\lag_\mu$ vanishes:
    \begin{gather*}
 i_{\xi_P}\left(d(\langle \mu +\delta,\A^P\rangle)+\B\right)   = {\cal L}_{\xi_P}(\langle \mu +\delta,\A^P\rangle)- d(\langle\mu+\delta,\xi\rangle) + d\delta_{\xi}  \\
 \phantom{i_{\xi_P}\left(d(\langle \mu +\delta,\A^P\rangle)+\B\right)  }{}
 = \langle \xi_P(\mu+\delta),\A^P\rangle + \langle\mu+\delta,{\rm ad}_{-\xi}\cdot\A^P\rangle - d\delta_\xi  + d\delta_{\xi}\\
 \phantom{i_{\xi_P}\left(d(\langle \mu +\delta,\A^P\rangle)+\B\right)  }{}
  = \langle \xi_P(\delta)-{\rm ad}^*_\xi\mu-{\rm ad}^*_\xi \delta,\A^P\rangle =0,
\end{gather*}
where we used the fact that $\xi\in\lag_\mu$, or equivalently ${\rm ad}^*_\xi\mu = i_\xi\Sigma_\delta$.

 5. Similar to cotangent bundle reduction, we use a `shift map' to construct the required dif\/feomorphism. Recall that $V\pi^Q$ denotes the bundle of tangent vectors vertical to $\pi^Q:Q\to Q/G$. The subbundle $V^0\pi^Q$ of $T^*Q$ is def\/ined as the annihilator of $V\pi^Q$. Below we introduce a~{\em shift map} $\phi_\mu^{\A}$ between $J^{-1}_\delta(\mu)$ and $V^0_P\pi^Q = V^0\pi^Q\times_Q P$ and we show that it is equivariant w.r.t.\ to actions of $G_\mu$ obtained by restriction of $\Psi^{T_PQ}$ to $J_\delta^{-1}(\mu)$ and $V^0_P\pi^Q$. This is suf\/f\/icient for~$\phi_\mu^{\A}$ to project to a dif\/feomorphism $[\phi_\mu^{\A}]$ between the quotient spaces. This provides us with the desired dif\/feomorphism because the quotient of $V^0_P\pi^Q$ is well known: $V^0_P\pi^Q/G_\mu = T^*_{P/G_\mu}(Q/G)$.

{\bf Def\/inition of the shift map.} In the following we use a slight abuse of notations: if we write $\langle \mu+\delta(p),\A^Q(q)\rangle$ for some $p\in P$ and $q=\e(p)$, then this is the cotangent vector in~$T_q^*Q$ determined by $v_q\mapsto \langle \mu+\delta(p),\A^Q(q)(v_q)\rangle$. Let $\phi_\mu^{\A}$ be the map
\[\phi_\mu^\A:\ J_\delta^{-1}(\mu) \to V^0_P\pi_Q; \qquad (\alpha_q,p) \mapsto \big(\alpha_q-\langle \mu+\delta(p),\A^Q(q)\rangle,p\big).\] It is well def\/ined, i.e.\ $\phi^\A_\mu(\alpha_q,p)\in V^0_P\pi^Q$ for any $(\alpha_q,p)$ in the level set $J_\delta^{-1}(\mu)$ since
\begin{gather*}
\langle \phi_\mu^{\A}(\alpha_q,p),(\xi_Q(q),p)\rangle   = \langle \alpha_q, \xi_Q(q)\rangle - \langle \mu+\delta(p),\A^Q(q)(\xi_Q(q))\rangle \\
 \phantom{\langle \phi_\mu^{\A}(\alpha_q,p),(\xi_Q(q),p)\rangle}= \langle J_\delta(\alpha_q,p),\xi\rangle - \langle \mu,\xi\rangle =0.
 \end{gather*}

{\bf Equivariance of the shift map.} For arbitrary $g\in G_\mu$, \begin{gather*}
\phi_\mu^{\A}\big(\Psi^{T_P^*Q}_g(\alpha_q,p)\big)  = \left(T^*\Psi^{Q}_{g^{-1}}(\alpha_q) - \langle \mu+\delta(pg) , \A^Q (qg) \rangle ,pg\right)\\
\phantom{\phi_\mu^{\A}(\Psi^{T_P^*Q}_g(\alpha_q,p))}{}
=  \left(T^*\Psi^{Q}_{g^{-1}}\left(\alpha_q - \langle \mu+\delta(pg) , {\rm Ad}_{g^{-1}}\cdot \A^Q(q)\rangle\right),pg\right)\\
\phantom{\phi_\mu^{\A}(\Psi^{T_P^*Q}_g(\alpha_q,p))}{}
= \left( T^*\Psi^{Q}_{g^{-1}}\left(\alpha_q - \langle \mu+\delta(p),\A^Q(q)\rangle\right),pg\right) = \Psi^{T_P^*Q}_g\left(\phi^\A_\mu(\alpha_q,p)\right).
\end{gather*}The projection of $\phi_\mu^{\A}$ is thus well def\/ined, and it is denoted by $[\phi_\mu^\A]$, i.e.\ $[\phi_\mu^\A]:J^{-1}_\delta(\mu)/G_\mu \to V^0_P \pi^Q/G_\mu=T^*_{P/G_\mu}(Q/G)$.
 \end{proof}

\begin{definition} $\tilde \B$ is the 2-form on $P/G_\mu$ obtained after reducing $\B + d(\langle \mu+\delta,\A^P\rangle)$.
\end{definition}
In the following proposition we put a symplectic structure on the f\/ibred product $T^*_{P/G_\mu}(Q/G)$ $=T^*(Q/G)\times_{Q/G} P/G_\mu$. The notations are similar to those on $T^*_PQ$.
\begin{definition}
We def\/ine the projections
\begin{enumerate}\itemsep=0pt
\item[1)] $\tilde\e_1:T_{P/G_\mu}^*(Q/G) \to T^*(Q/G)$,
\item[2)] $\tilde\e_2: T_{P/G_\mu}^*(Q/G) \to P/G_\mu$.
\end{enumerate} The canonical symplectic 2-form on $T^*(Q/G)$ is denoted by $\omega_{Q/G}$.\end{definition}
Completely analogous to the construction on $T^*_PQ$, we introduce $\tilde\e_1^*\omega_{Q/G} +\e^*_2\tilde\B$ as a  2-form on $T^*_{P/G_\mu}(Q/G)$.
\begin{figure}[thb]
\centering
\includegraphics{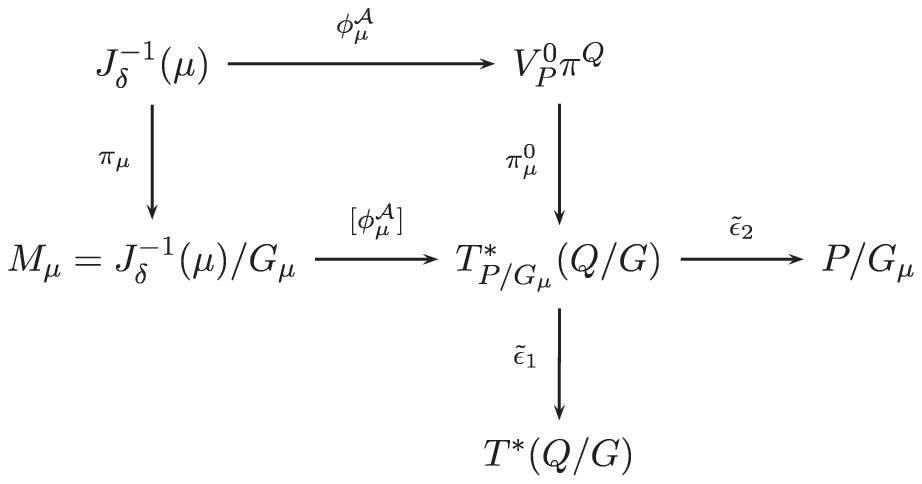}

\caption{Cotangent bundle reduction (notations as in Theorem~\ref{thm:mw} with $M=T^*_PQ$).}\end{figure}

\begin{theorem}[Generalized magnetic cotangent bundle reduction]
  Let $\mu$ denote a regular value of the momentum map $J_\delta$ for the symplectic structure $(T^*_PQ,\e_1^*\omega_Q+\e^*_2\B)$ with isotropy subgroup $G_\mu$, and let $\A^Q$ be a principal connection on $Q\to Q/G$. Then the symplectically reduced manifold $(M_\mu,\omega_\mu)$ is symplectomorphic to $(T^*_{P/G_\mu}(Q/G),\tilde\e_1^*\omega_{Q/G} + \tilde\e_2^*\tilde\B))$, with symplectomorphism $[\phi^{\A}_\mu]$.
\end{theorem}

Although this result is a straightforward extension of standard magnetic cotangent bundle reduction in~\cite{MarsdenHamRed}, we include a proof because its method turns out to be useful later on.
\begin{proof}
Let $i_0$ denote the injection $V^0_P\pi^Q\to T^*_PQ$ and $\pi_\mu^0$ the projection $V^0_P\pi^Q\to T^*_{P/G_\mu}(Q/G)$. We know that $[\phi_\mu^\A]$ is a dif\/feomorphism, and therefore it only remains to show that the 2-form $\tilde\e_1^*\omega_{Q/G} + \tilde\e_2^*\tilde\B$ is pull-backed to $\omega_\mu$ under this map (and therefore, implicitly $\tilde\e_1^*\omega_{Q/G} + \tilde\e_2^*\tilde\B$ will be nondegenerate).

We use the fact that $\omega_\mu$ is uniquely determined by $i^*_\mu\omega_Q=\pi^*_\mu\omega_\mu$, with $i_\mu:J^{-1}_\delta(\mu)\to T_P^*Q$ the natural inclusion and $\pi_\mu:J^{-1}_\delta(\mu)\to M_\mu$ the projection to the quotient space. Due to the uniqueness property, it is therefore suf\/f\/icient to show that \begin{gather}\label{eq:symplmorph}
\pi^*_\mu\big([\phi_\mu^\A]^*(\tilde\e_1^*\omega_{Q/G} + \tilde\e_2^*\tilde\B)\big)=i^*_\mu(\e_1^*\omega_Q+\e^*_2\B).
\end{gather}

{\bf The left-hand side of~\eqref{eq:symplmorph}.} Since $[\phi_\mu^\A]\circ\pi_\mu = \pi^0_\mu\circ\phi_\mu^\A$ the left-hand side of the above equation is $(\pi^0_\mu\circ\phi_\mu^\A)^*\big(\tilde\e_1^*\omega_{Q/G} + \tilde\e_2^*\tilde\B\big)$.

{\bf The right-hand side of~\eqref{eq:symplmorph}.}
We f\/irst show the equality of two 1-forms:
\[
i^*_\mu(\e_1^*\theta_Q)\qquad \mbox{and} \qquad \big(\phi_\mu^\A\big)^*\Big(i^*_0\big(\e_1^*\theta_Q +\e_2^*\langle \mu+\delta,\A^P\rangle\big)\Big).
\] Let $(\alpha_q,p)\in J^{-1}_\delta(\mu)$, $V_{(\alpha_q,p)}\in T( T_P^*Q)$ a tangent vector to $J^{-1}_\delta(\mu)$ and $v_q\in T_qQ$ denotes the projection of $V_{(\alpha_q,p)}$. Then, the f\/irst 1-form equals
\[
i^*_\mu(\e_1^*\theta_Q)(\alpha_q,p)(V_{(\alpha_q,p)})= \theta_Q(\alpha_q)(T\e_1(V_{(\alpha_q,p)}))=\langle \alpha_q, v_q\rangle.
\]
The second 1-form evaluated on this tangent vector gives
\begin{gather*}
   (\phi_\mu^\A)^*\Big(i^*_0\big(\e_1^*\theta_Q +\e_2^*\langle\mu+\delta,\A^P\rangle\big)\Big)(\alpha_q,p)(V_{(\alpha_q,p)}) \\
    \qquad{}  = \theta_Q\big(\alpha_q-\langle\mu+\delta(p),\A^Q(q)\rangle\big)\big(T\e_1(T(\phi^\A_\mu)(V_{(\alpha_q,p)}))\big)  + \langle \mu+\delta(p),\A^Q(q)(v_q)\rangle\\
{} \qquad = \langle \alpha_q, v_q\rangle.
\end{gather*}
The two 1-forms are identical and after taking the exterior derivative and adding $\e^*_2\B$, we obtain that the right-hand side of~\eqref{eq:symplmorph} equals \[i^*_\mu(\e_1^*\omega_Q+\e^*_2\B) = \big(\phi_\mu^{\A}\big)^*\Big(i^*_0\big(\e_1^*\omega_Q +\e_2^*(d\langle \mu+\delta,\A^P\rangle+\B)\big)\Big).\]
Since (i) $(\tilde\e_2\circ\pi^0_\mu)^*\tilde\B = (\e_2\circ i_0)^*(d\langle \mu+\delta,\A^P\rangle+\B)$, and (ii) $(\tilde\e_1\circ\pi^0_\mu)^*\theta_{Q/G}=i_0^*\e_1^*\theta_{Q}$, one easily verif\/ies that the left-hand side equals the right-hand side in~\eqref{eq:symplmorph}.
\end{proof}

We conclude with a result on the behavior of a symplectomorphism under symplectic reduction. Assume two symplectic manifolds $(M,\Omega)$ and $(M',\Omega')$ and a symplectomorphism $f:M\to M'$ (i.e.\ a map for which $f^*\Omega'=\Omega$) are given. We assume in addition that both $M$ and $M'$ are equipped with a canonical free and proper action of $G$. Let $J:M\to \lag^*$ and $J':M'\to \lag^*$ denote corresponding momentum maps for these actions on $M$ and $M'$ respectively. We say that $f$ is equivariant if $f(mg)=f(m)g$ for arbitrary $m\in P$, $g\in G$. Note that the non-equivariance cocycles for $J$ and $J'$ are equal up to a coboundary. Without loss of generality we may assume that $f^*J'=J$ and that the non-equivariance cocycles coincide. This in turn guarantees that the af\/f\/ine actions on $\lag^*$ coincide and that the isotropy groups of an element $\mu\in\lag^*$ coincide for both af\/f\/ine actions. Finally, f\/ix  a regular value $\mu \in \lag^*$ for both $J$ and $J'$.
\begin{theorem}\label{thm:1}
  If $f$ is an equivariant symplectic diffeomorphism $M\to M'$, such that $J'=J \circ f$, then under symplectic reduction, the symplectic manifolds $(M_\mu,\Omega_\mu)$ and $(M'_\mu,\Omega'_\mu)$ are symplectically diffeomorphic under the map
  \[
  [f_\mu]:\ M_\mu\to M_\mu'; \qquad [m]_{G_\mu} \mapsto [f(m)]_{G_\mu}.
  \]
\end{theorem}

\begin{proof}
  This is a straightforward result. Since $f$ is a dif\/feomorphism for which \mbox{$J'=J\circ f$}, the restriction $f_\mu$ of $f$ to $J^{-1}(\mu)$ determines a dif\/feomorphism from $J^{-1}(\mu)$ to $J'^{-1}(\mu)$. The equivariance implies that $f_\mu$ reduces to a dif\/feomorphism $[f_\mu]$ from $M_\mu=J^{-1}(\mu)/G_\mu$ to $M'_\mu=J'^{-1}(\mu)/G_\mu$.
 It is our purpose to show that $[f_\mu]^*\Omega'_\mu = \Omega_\mu$, or since both $\pi_\mu$ and $\pi'_\mu$ are projections, that $\pi_\mu^*\Omega_\mu=f^*_\mu(\pi'^*_\mu\Omega'_\mu)$. The determining property for $\Omega_\mu$ is $\pi^*_\mu\Omega_\mu = i^*_\mu\Omega$ (and similarly for $\Omega'_\mu$). From diagram chasing we have that $i^*_\mu\Omega = f_\mu^*(i'^*_\mu \Omega')$. Then \[\pi_\mu^*\Omega_\mu = i^*_\mu\Omega=f^*_\mu(i'^*_\mu \Omega') = f^*_\mu(\pi'^*_\mu \Omega_\mu') =\pi^*_\mu([f_\mu]^*\Omega_\mu'),\]
  since $\pi'_\mu\circ f_\mu = [f_\mu]\circ\pi_\mu$ by def\/inition. This concludes the proof.
\end{proof}

\subsection{Routh reduction for magnetic Lagrangian systems}\label{section5.2}

In Proposition~\ref{prop:equiv} we have introduced the momentum map $J_{L,\delta} = J_\delta \circ \F L$ for a $G$-invariant hyperregular magnetic Lagrangian system $(\e:P\to Q, L,\B)$, i.e.\ \[\langle J_{L,\delta}(v_q,p),\xi\rangle = \langle \F L(v_q,p),(\xi_Q(q),p)\rangle - \delta_\xi(p).\] We know from Proposition~\ref{prop:equivleg} that the Legendre transform $\F L$ is equivariant. After restriction to $J_{L,\delta}^{-1}(\mu)$ it reduces to a symplectic dif\/feomorphism between the symplectically reduced spaces $J_{L,\delta}^{-1}(\mu)/G_\mu$ and $J_{\delta}^{-1}(\mu)/G_\mu$ (see also Theorem~\ref{thm:1}). The following diagram summarizes these previous observations.
\begin{figure}[htb]
\centering
\includegraphics{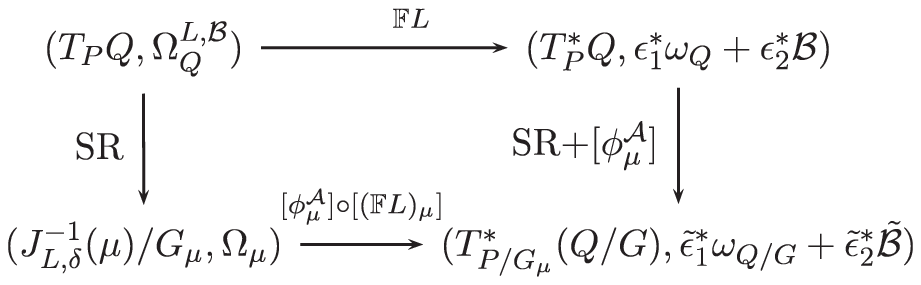}

\caption{Diagram relating tangent and cotangent reduction (SR = symplectic reduction).}\label{fig:diag2}
\end{figure}

\begin{figure}[htb]
\centering
\includegraphics{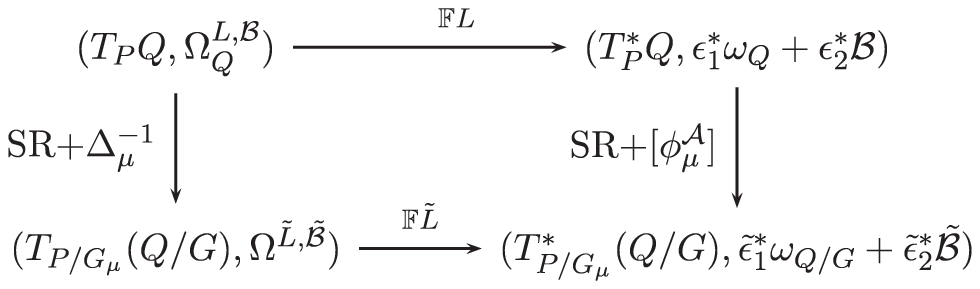}

\caption{A symplectic view on Routh reduction.}\label{fig:diag3}\end{figure}

We are now only three steps away from a description of Routh reduction for magnetic Lagrangian systems. The f\/inal goal is to describe the symplectic reduced manifold $(J_{L,\delta}^{-1}(\mu)/G_\mu,\Omega_\mu)$ as a symplectic manifold associated with a new magnetic Lagrangian system, with Lagrangian say $L_\mu$.
\begin{enumerate}\itemsep=0pt \item The f\/irst step is the construction of a dif\/feomorphism $\Delta_\mu:T_{P/G_\mu}(Q/G)\to J_{L,\delta}^{-1}(\mu)/G_\mu$.
\end{enumerate} This is crucial because the Hamiltonian dynamics determined by the Euler--Lagrange equations reduce to Hamiltonian dynamics on a manifold of the form $T_{P/G_\mu}(Q/G)$. If we can characterize this reduced dynamics as the Euler--Lagrange equations of a magnetic Lagrangian system with conf\/iguration manifold $P/G_\mu \to Q/G$, we have developed a Routh reduction technique for magnetic Lagrangian systems. This characterization consists of the two remaining steps mentioned before:
\begin{enumerate}\setcounter{enumi}{1}\itemsep=0pt \item We show that the composition $[\phi_\mu^{\A}]\circ [(\F L)_\mu] \circ \Delta_\mu$ equals the f\/ibre derivative of some Lagrangian function $\tilde L$ on $T_{P/G_\mu}(Q/G)$, and this implies that the symplectic 2-form $\Delta^*_\mu\Omega_\mu$ on $T_{P/G_\mu}(Q/G)$ is of the form
\[
\F \tilde L^*(\tilde\e_1^*\omega_{Q/G} + \tilde\e_2^*\tilde\B),
\]
which is the symplectic structure associated to the {\em reduced} magnetic Lagrangian system $(P/G_\mu \to Q/G, \tilde L, \tilde\B)$;
\item We show that the reduction of the energy $E_L$ in the symplectic reduction scheme is precisely the energy $E_{\tilde L}$ on $T_{P/G_\mu}(Q/G)$.  This guarantees that the Euler--Lagrange equations associated to the reduced Lagrangian $\tilde L$ are related to the Euler--Lagrange equations for~$L$.
\end{enumerate} The new reduced Lagrangian $\tilde L$ is what we call the Routhian (it is often alternatively denoted by $R$ or $R^\mu$). This is summarized in the diagram in Fig.~\ref{fig:diag3}.

{\bf Step~1: the def\/inition of $\Delta_\mu: T_{P/G_\mu}(Q/G)\to J^{-1}_{L,\delta}(\mu)/G_\mu$.}
First consider the map $\Pi_\mu:J^{-1}_{L,\delta}(\mu)/G_\mu\to T_{P/G_\mu} (Q/G)$ which is obtained from the $G_\mu$-invariance of the map \[J^{-1}_{L,\delta}(\mu)\to T_{P/G_\mu}(Q/G); (v_q,p) \mapsto (T\pi^Q(v_q),[p]_{G_\mu}).\] We will def\/ine the map $\Delta_\mu$ as the inverse of  $\Pi_\mu$. In general however, $\Pi_\mu$ is not invertible. The following def\/inition is the analogue of $G$-regularity for a classical Lagrangian.

\begin{definition} The Lagrangian $L$ of a $G$-invariant magnetic Lagrangian system is called $G$-regular if the map $J_{L,\delta}|_{(v_q,p)}: \lag \to \lag^*; \xi \mapsto J_{L,\delta}(v_q +\xi_Q(q),p) $ is a dif\/feomorphism for all $(v_q,p)\in T_PQ$.
\end{definition}
Every mechanical $G$-invariant magnetic Lagrangian system is $G$-regular because $J_{L,\delta}|_{(v_q,p)}$ is an af\/f\/ine map modeled on the metric on $\lag$ obtained from the kinetic energy metric.

\begin{proposition}\label{prop:gregular} $\Pi_\mu$ is a diffeomorphism if the Lagrangian is $G$-regular.  \end{proposition}
\begin{proof}
We construct the inverse for $\Pi_\mu$. Choose an element $(v_{[q]_G},[p]_{G_\mu})$ in $T_{P/G_\mu}(Q/G)$ and f\/ix a point $(v_q,p)\in T_PQ$ that projects onto $(v_{[q]_G},[p]_{G_\mu})$, i.e.\ $T\pi^Q(v_q)=v_{[q]}$ and $p\in[p]_{G_\mu}$. Due to the $G$-regularity of $L$, there exists a unique element $\xi$ in $\lag$ such that $J_{L,\delta}(v_q +\xi_Q(q),p) = \mu$. The orbit of the point $(v_q+\xi_Q(q),p)$ under the $G_\mu$-action determines an element in $J_{L,\delta}^{-1}(\mu)/G_\mu$. We will show that this construction is independent of the chosen point $(v_q,p)$, and therefore def\/ines a map $\Delta_\mu: T_{P/G_\mu} (Q/G)\to J^{-1}_{L,\delta}(\mu)/G_\mu$ which is the inverse to $\Pi_\mu$.

To show that the construction is independent of the chosen point $(v_q,p)$ in $T_PQ$, we choose any other point $(v'_q,p')$ in $T_PQ$ projecting onto $(v_{[q]_G},[p]_{G_\mu})$. The point is of the form $(v'_q,p')=(\Psi^{TQ}_g(v_q+\eta_Q(q)),pg)$ for some $g\in G_\mu$ and $\eta\in\lag$. We now repeat the previous  construction applied to $(v'_q,p')$: we consider a unique element $\xi'\in\lag$ for which $J_{L,\delta}(\Psi^{TQ}_g(v_q+\eta_Q(q)) +\xi'_Q(qg),pg) = \mu$. Due to the equivariance of $J_{L,\delta}$, we conclude that
\begin{gather*}
\mu = J_{L,\delta}(\Psi^{TQ}_g(v_q+\eta_Q(q)) +\xi'_Q(qg),pg) ={\rm Ad}_g^*\cdot J_{L,\delta}(v_q+(\eta+{\rm Ad}_g\xi')_Q(q),p) -\sigma_\delta(g^{-1}).
 \end{gather*}
 Since $g\in G_\mu$, we may conclude that $J_{L,\delta}(v_q+(\eta+{\rm Ad}_g\xi')_Q(q),p)=\mu$ and therefore $\eta+{\rm Ad}_g\xi' = \xi$. This implies that $(v'_q+\xi'_Q(q),p')=\Psi^{T_PQ}_g(v_q+\xi_Q(q),p)$ for $g\in G_\mu$ such that $p'=pg$. This concludes the proof.
\end{proof}

The previous proposition guarantees that for a given $G$-regular Lagrangian and a bundle adapted coordinate chart $(x^r,g^m)$ in $Q\to Q/G$ and $(x^r,g^m,p^a)$ in $P\to Q$, the functions $(x^r,\dot x^r,g^m,p^a)$ determine a coordinate chart in $J^{-1}_{L,\delta}(\mu)$. This is useful in the following.

{\bf Step~2: the reduced symplectomorphism.}
We will now introduce a new Lagrangian $\tilde L$ on $T_{P/G_\mu}(Q/G)$ with the property that its Legendre transform $\F \tilde L: T_{P/G_\mu} (Q/G) \to T^*_{P/G_\mu}(Q/G)$ is precisely the reduced Legendre transform $[(\F L)_\mu]$, more specif\/ically $\F \tilde L= [\phi^\A_\mu]\circ[(\F L)_\mu] \circ \Delta_\mu$.

As is already described in~\cite{BC}, the structure of the Lagrangian $\tilde L$ is completely determined by the cotangent bundle reduction scheme, and in particular by the shift map $\phi^\A_\mu$ from $J^{-1}_\delta(\mu) \to V^0_P\pi^Q$. To make this statement more precise, we remark that $[\phi^\A_\mu]\circ[(\F L)_\mu]$ is the quotient map of $\phi^\A_\mu\circ \F L$. Thus, if $[(v_q,p)]_{G_\mu}$ is arbitrary in $J^{-1}_{L,\delta}(\mu)/G_\mu$, then $[\phi^\A_\mu]\circ[(\F L)_\mu]([(v_q,p)]_{G_\mu})$ is the quotient of \[\F L(v_q,p) - \left(\langle\mu+\delta(p),\A^Q(q)\rangle,p\right).\]
 The latter is the f\/ibre derivative of a new function on $T_PQ$, namely $L(v_q,p) {-} \langle \mu{+}\delta(p),\A^Q(q)(v_q)\rangle.\!$ This function is clearly $G_\mu$-invariant.

\begin{definition} The function $\tilde L$ is def\/ined as the pull-back under $\Delta_\mu$ of the quotient map of the restriction to $J^{-1}_{L,\delta}(\mu)$ of the function \[
(v_q,p) \mapsto L(v_q,p)- \left\langle \mu+\delta(p), \A^Q(q)(v_q)\right\rangle.
\]
\end{definition}

\begin{lemma}\label{lem:legendre}
  $\F \tilde L=[\phi^\A_\mu]\circ[(\F L)_\mu]\circ\Delta_\mu$.
\end{lemma}
\begin{proof}
Fix elements $(v_{[q]_G},[p]_{G_\mu})\in T_{P/G_\mu}(Q/G)$ and f\/ix a representative $(v_q,p)\in J^{-1}_{L,\delta}(\mu)$ in the orbit $\Delta_\mu(v_{[q]_G},[p]_{G_\mu})\in J^{-1}_{L,\delta}(\mu)/G_\mu$. By def\/inition of the maps involved, we have
\begin{gather*}
\big([\phi^\A_\mu]\circ[(\F L)_\mu]\circ\Delta_\mu\big)(v_{[q]_G},[p]_{G_\mu})   = (\pi^0_\mu \circ \phi^\A_\mu)(\F L(v_q,p))\\
 \phantom{\big([\phi^\A_\mu]\circ[(\F L)_\mu]\circ\Delta_\mu\big)(v_{[q]_G},[p]_{G_\mu})}{}
 = \pi^0_\mu\big(\F L(v_q,p)-\big(\langle \mu+\delta(p),\A^Q(q)\rangle,p\big)\big).
\end{gather*}
Next we study the f\/ibre derivative of the map $\tilde L$.  Fix a point $(w_{[q]_G},[p]_{G_\mu})$ and we compute
\[
\left\langle \F\tilde L (v_{[q]_G},[p]_{G_\mu}) , (w_{[q]_G},[p]_{G_\mu})\right\rangle=\left. \frac{d}{du}\right|_{u=0}\tilde L (v_{[q]_G}+u w_{[q]_G},[p]_{G_\mu}).
\]
We construct a curve $u\mapsto \zeta(u)$ in $J^{-1}_{L,\delta}(\mu)$ that projects onto the curve $u\mapsto\Delta_\mu(v_{[q]_G}+u w_{[q]_G},[p]_{G_\mu})$ in $J^{-1}_{L,\delta}(\mu)/G_\mu$ such that $\zeta(0)=(v_q,p)$ and $\dot \zeta(0)$ is vertical to the projection $\rho_2\circ i_\mu:J^{-1}_{L,\delta}(\mu)\to P$ (recall that $\rho_2: T_PQ\to P$ is the projection onto the second factor in the f\/ibred product).

The existence of such a curve is best shown in a specif\/ic coordinate chart (see above) for $J^{-1}_L(\mu)$: $(x^r,\dot x^r, g^m,p^a)$  where the index $r=1,\ldots,\dim Q/G$, $m=1,\ldots,\dim G$ and $a=1,\ldots ,\dim P-\dim Q$. Note that $(x^r,\dot x^r)$ is a coordinate chart in $T(Q/G)$.
In these coordinates, we let $(v_q,p)=(x^r_0,\dot x^r_0,g^m_0,p^a_0)$ and $w_{[q]_G}=(x^r_0, w^r_0)$. We def\/ine the curve $\zeta(u)$ to be the curve $u\mapsto (x^r_0,\dot x^r_0+u w^i_0,g^m_0,p^a_0)$. Then the tangent vector to $T\rho_1(\dot \zeta(0))$ is the vertical lift of some $w_q\in T_qQ$ with $T\pi^Q(w_q)=w_{[q]_G}$.

Finally, from the def\/inition of $\tilde L$ and the f\/ibre derivative $\F\tilde L$ we obtain
\begin{gather*}
\left\langle \F\tilde L(v_{[q]_G},[p]_{G_\mu}) , (w_{[q]_G},[p]_{G_\mu})\right\rangle
=\left.\frac{d}{du}\right|_{u=0}\left(L - \big\langle\mu+\delta, \A^Q\big\rangle\right)(\zeta(u))\\
\phantom{\left\langle \F\tilde L(v_{[q]_G},[p]_{G_\mu}) , (w_{[q]_G},[p]_{G_\mu})\right\rangle}{}
= \langle \F L(v_q,p)-(\langle\mu+\delta(p),\A^Q(q) \rangle, p),(w_q,p)\rangle\\[2mm]
\phantom{\left\langle \F\tilde L(v_{[q]_G},[p]_{G_\mu}) , (w_{[q]_G},[p]_{G_\mu})\right\rangle}{}
= \left\langle \pi^0_\mu\big(\F L(v_q,p)-\big(\langle \mu+\delta(p),\A^Q(q)\rangle,p\big)\big), (w_{[q]_G},[p]_{G_\mu})\right\rangle,
\end{gather*}
since $\F L(v_q,p)-(\langle \mu+\delta(p),\A^Q(q)\rangle,p) \in V^0_P\pi^Q$. This  concludes the proof.
\end{proof}

{\bf Step~3: the reduced energy Hamiltonian.}
The third and last step concerns the specif\/ic reduced dynamics. We have to relate the energy $E_L$  on $T_PQ$ to the energy of the Routhian $\tilde L$ on $T_{P/G_\mu}(Q/G)$.   In the following Lemma we again use the notations from Theorem~\ref{thm:mw} applied to the reduction of the symplectic structure on $(T_PQ,\Omega^{L,\B})$.
\begin{lemma}\label{lem:energy}
  The energy $E_{\tilde L}$ is the reduced Hamiltonian, i.e.\ it satisfies:
  \[
  (\Pi_\mu\circ\pi_\mu)^*E_{\tilde L} = i^*_\mu E_L,
  \]
  with $\Pi_\mu\circ\pi_\mu: J^{-1}_{L,\delta}(\mu)\to J^{-1}_{L,\delta}(\mu)/G_\mu\to T_{P/G_\mu}(Q/G)$ and $i_\mu:J^{-1}_{L,\delta}(\mu)\to T_PQ$.
\end{lemma}
\begin{proof}
Let $(v_q,p)\in J^{-1}_{L,\delta}(\mu)$, such that $(\Pi_\mu\circ\pi_\mu)(v_q,p)=(v_{[q]_G},[p]_{G_\mu})$. Then
\begin{gather*}
  i^*_\mu E_L(v_q,p) = \langle\F L(v_q,p),(v_q,p)\rangle-L(v_q,p)\\
  \phantom{i^*_\mu E_L(v_q,p)}{}
  = \langle  (\phi^\A_\mu\circ(\F L)_\mu)(v_q,p),(v_q,p)\rangle - \left(L(v_q,p)-\langle \mu+\delta(p),\A^Q(q)(v_q)\rangle\right)\\
 \phantom{i^*_\mu E_L(v_q,p)}{}
 = \left\langle \big([\phi^\A_\mu]\circ[(\F L)_\mu]\circ\Delta_\mu\big)(v_{[q]_G},[p]_{G_\mu}),(v_{[q]_G},[p]_{G_\mu})\right\rangle - \tilde L (v_{[q]_G},[p]_{G_\mu}).
\end{gather*}
Using the result from Lemma~\ref{lem:legendre} this concludes the proof.
\end{proof}

{\bf Routh reduction.} The previous three steps are summarized in the following theorem.
\begin{theorem}[Routh reduction for magnetic Lagrangian systems]\label{thm:routh}
  Let $(\e:P\to Q,L,\B)$ be a~hyperregular, $G$-invariant and $G$-regular magnetic Lagrangian system and let $\delta$ be a~$\B\lag$-potential of the magnetic term $\B$ with $1$-cocycle $\sigma_\delta$.
  \begin{enumerate}\itemsep=0pt
  \item[$1.$] Let $\mu\in\lag^*$ be a regular value of the momentum map $J_{L,\delta}$ and let $G_\mu$ be the isotropy subgroup of $\mu$ w.r.t.\ the affine action on $\lag^*$ with $1$-cocycle $-\sigma_\delta$, i.e.\ $g\in G_\mu$ if and only if $\mu = {\rm Ad}^*_g\mu -\sigma_\delta(g^{-1})$.
  \item[$2.$] Fix a connection $\A^Q$ on $\pi^Q:Q\to Q/G$ and let $\A^P$ be the corresponding connection on $P\to P/G$. Compute the restriction of the $G_\mu$-invariant function $L(v_q,p)-\langle \mu+\delta(p),\A^Q(q)(v_q)\rangle$ to  $J^{-1}_{L,\delta}(\mu)$ and let $\tilde L$ be its quotient  to $T_{P/G_\mu}(Q/G) \cong J^{-1}_{L,\delta}(\mu)/G_\mu$.
 \item[$3.$] Compute $\tilde\B$ as the projection to $P/G_\mu$ of the $2$-form $\B+d\langle \mu+\delta,\A^P\rangle$.
\item[$4.$]  Consider the magnetic Lagrangian system: $(\tilde\e_\mu:P/G_\mu \to Q/G, \tilde L, \tilde \B)$.
 \end{enumerate}

 This reduced magnetic Lagrangian system is hyperregular and every solution $p(t)\in P$ to the Euler--Lagrange equations for $(\e,L,\B)$ with momentum $\mu$ projects under $P\to P/G_\mu$ to a solution of the Euler--Lagrange equations for $(\tilde\e_\mu:P/G_\mu \to Q/G, \tilde L, \tilde B)$. Conversely, every solution to the Euler--Lagrange equations for $(\tilde\e_\mu:P/G_\mu \to Q/G,\tilde L, \tilde B)$ is the projection of a solution to the Euler--Lagrange equations for $(\e,L,\B)$ with momentum $\mu$.
\end{theorem}

It is possible to say more about the structure of $\tilde \B$ and its relation to the connection $\A^Q$.  This and reconstruction aspects fall out of the scope of this paper. We refer to~\cite{BM} where these topics are described in more detail. They carry over to this more general framework in a~straightforward way.

\subsection{Reduction of magnetic Lagrangian systems on Lie groups}
\label{sec:liegroups}

Consider a magnetic Lagrangian system on $P=Q=G$, i.e.\ the conf\/iguration space is a Lie group $G$. We start from a function $\ell$ on $\lag$ and with it we associate a Lagrangian $L$ on $TG$ by right multiplication $L(g,v_g)=\ell(v_g g^{-1})$. By def\/inition $L$ is invariant under the right action of $G$ on itself. We assume that a magnetic 2-form $\B$ is given which is invariant under right multiplication and admits a $\B\lag$-potential $\delta:G\to \lag^*$. We f\/irst rephrase some def\/initions in this specif\/ic setting.

\begin{enumerate}\itemsep=0pt
\item The 1-cocycle $\sigma_\delta: G\to \lag^*$ satisf\/ies $\sigma_\delta(g) = \delta(hg^{-1})- {\rm Ad}^*_{g^{-1}} \delta(h)$, for arbitrary $h\in G$. If we let $h=e$, then $\sigma_\delta(g) = \delta(g^{-1}) - {\rm Ad}^*_{g^{-1}} \delta(e)$ or equivalently, ${\rm Ad}^*_{g}\sigma_\delta(g) ={\rm Ad}^*_{g} \delta(g^{-1}) - \delta(e)$. Similarly if we let $h=g$, then $\sigma_\delta(g) = \delta(e) - {\rm Ad}^*_{g^{-1}} \delta(g)$. Since $\delta$ is determined up to a~constant, we may assume without loss of generality that $\delta(e)=0$.
\item The associated 2-cocycle $\Sigma_\delta(\xi,\eta) =  \xi_G(\delta_\eta) - \delta_{[\xi,\eta]}$.
\item We use the right identif\/ication of $TG$ with $G\times\lag$, i.e.\ $(g,v_g)$ is mapped to $(g,v_gg^{-1})\in G\times\lag$. The right action of $G$ on $TG$ equals right multiplication in the f\/irst factor of $G\times\lag$ under this identif\/ication.
\item We use the Maurer--Cartan principal connection on $G\to G/G$: $\A(g)(v_g) = g^{-1} v_g$. In the right identif\/ication, the connection corresponds to the map $(g,\xi)\in G\times\lag \mapsto {\rm Ad}_{g^{-1}} \xi\in\lag$.
\item The momentum map $J_{L,\delta}:TG\to \lag^*$ equals $J_{L,\delta}(g,\xi g) = {\rm Ad}^*_{g} \F \ell (\xi)-\delta(g)$, for $(g,\xi)\in G\times \lag$ arbitrary. If $L$ is $G$-regular then $\F\ell$ is invertible, i.e.\ there exists a function $\chi: \lag^*\to \lag$ such that $\F\ell( \chi (\nu)) = \nu$.
\item The af\/f\/ine action on $\lag^*$ is $(g,\mu)\mapsto {\rm Ad}^*_{g} \mu - \sigma_\delta(g^{-1})$.
\item The isotropy group $G_\mu$ consists of group elements $g$ such that $\mu +\delta(g) = {\rm Ad}^*_g\mu$.
\item The quotient $G/G_\mu$ (right coset space) can be identif\/ied with $\tilde\Or_\mu$, i.e.\ $[g]_{G_\mu}
 \in G/G_\mu$ is mapped onto $\nu={\rm Ad}_{g^{-1}}^*\mu - \sigma_\delta(g)={\rm Ad}^*_{g^{-1}}(\mu+\delta(g)) $. A tangent vector to $G/G_\mu$ at $[g]_{G_\mu}$ which is the projection of $\xi g$ is mapped to a vector $\dot \nu = -{\rm ad}_\xi^*\nu + i_\xi \Sigma_\delta$ in $\lag^*$.
\end{enumerate}

\begin{lemma} \label{lemma:sympform} The $2$-form $\B+d(\langle \mu +\delta, \A\rangle)$ reduces to the Kirillov--Kostant--Souriau symplectic $2$-form on $\tilde \Or_\mu \cong G/G_\mu$. \end{lemma}
\begin{proof}
Let $g$ be arbitrary, and let $v_g = \xi g$, $w_g= \eta g$ be two tangent vectors in $T_gG$ with $\xi,\eta\in\lag$ arbitrary. Note that $v_g=\xi'_G(g)$ with $\xi'= {\rm Ad}_{g^{-1}}\xi$, and similarly $w_g=\eta'_G(g)$ with $\eta'= {\rm Ad}_{g^{-1}}\eta$. Then \[\B(g)(v_g,w_g)  = \B(g)(\xi'_G(g),\eta'_G(g)) = \eta'_G(\delta_{\xi'})(g).\]  On the other hand
\[d(\langle \mu +\delta, \A\rangle)(g)(v_g,w_g)  = \xi'_G(\langle \mu+\delta, \eta'\rangle)(g)- \eta'_G(\langle \mu+\delta, \xi'\rangle)(g) - \langle\mu+\delta(g), [\xi',\eta']\rangle.\]
Before continuing, we compute the equivariance of the 2-cocycle $\Sigma_\delta(\xi',\eta')$:
\begin{gather*}
\Sigma_\delta(\xi',\eta')  = -\frac{d}{ds} \langle \sigma_\delta(g^{-1}\exp s  \xi g) ,\eta'\rangle
 =  -\frac{d}{ds} \langle \sigma_\delta(g^{-1}\exp s  \xi)  + {\rm Ad}^*_{\exp - s\xi g} \sigma_\delta(g),\eta'\rangle  \\
 \phantom{\Sigma_\delta(\xi',\eta')}{}
 =    -\frac{d}{ds} \langle \sigma_\delta(g^{-1}) + {\rm Ad}^*_{g} \sigma_\delta(\exp s\xi)  + {\rm Ad}^*_{g} {\rm Ad}^*_{\exp - s\xi} \sigma_\delta(g),{\rm Ad}_{g^{-1}}\eta\rangle \\
 \phantom{\Sigma_\delta(\xi',\eta')}{}
  = \Sigma_\delta(\xi,\eta) + \langle \sigma_\delta(g), [\xi,\eta]\rangle.
  \end{gather*}
Summarizing, we have
\begin{gather*}
\big(\B+d(\langle \mu +\delta, \A\rangle)\big)(g)(v_g,w_g)  =
\xi'_G(\langle \mu+\delta, \eta'\rangle)(g)  - \langle\mu+\delta(g), [\xi',\eta'] \rangle \\
 \phantom{\big(\B+d(\langle \mu +\delta, \A\rangle)\big)(g)(v_g,w_g)}{}
 = \Sigma_\delta(\xi',\eta') + \delta_{[\xi',\eta']} (g)  - \langle\mu+\delta(g), [\xi',\eta']\rangle \\
\phantom{\big(\B+d(\langle \mu +\delta, \A\rangle)\big)(g)(v_g,w_g)}{}
 = \Sigma_\delta(\xi,\eta) - \langle {\rm Ad}^*_{g^{-1}} \mu -\sigma_\delta(g) , [\xi,\eta] \rangle.
 \end{gather*}
This 2-form is reducible to a 2-form on $\tilde\Or_\mu$. If we use the isomorphism $G/G_\mu \to \tilde \Or_\mu$ introduced above, the 2-form reduces to $\tilde \B (\nu)(\dot \nu, \dot \nu') = \langle \dot \nu, \eta\rangle$ with $\eta\in\lag$ such that $\dot \nu ' = -{\rm ad}^*_\eta\nu + i_\eta\Sigma_\delta$.
\end{proof}
We conclude by computing the Routhian $\tilde L$ as a function on $\tilde \Or_\mu$. By def\/inition, it equals the reduction to $\tilde \Or_\mu$ of $L-\langle \mu+\delta,\A\rangle|_{J_{L,\delta} =\mu}$. In the right identif\/ication, the level set $J_{L,\delta}(g,v_g) =\mu$ is precisely ${\rm Ad}^*_{g^{-1}}(\mu +\delta(g)) = \F\ell(\xi)$, with $\xi g=v_g$. If we set $\nu = {\rm Ad}^*_{g^{-1}}(\mu +\delta(g)) \in \tilde\Or_\mu$, the f\/ixed momentum condition is $\xi = \chi(\nu)$. The Routhian  $\tilde L(\nu)$ becomes in the right identif\/ication \[\tilde L(\nu)=\ell(\chi(\nu)) - \langle \nu, \chi(\nu)\rangle.\] By application of the chain rule, it easily follows that $\langle d\tilde L (\nu), \dot \nu'\rangle  = -\langle \dot \nu',\chi(\nu)\rangle$.   The reduced Euler--Lagrange equations are
\begin{gather} \label{twiceredEL}
i_{\dot \nu} \tilde \B(\nu) = d \tilde L(\nu) \qquad \mbox{or} \qquad \dot \nu =- {\rm ad}^*_{\chi(\nu)}\nu + i_{\chi(\nu)} \Sigma_\delta.
\end{gather}
For later purpose, we remark that for a left action and $\ell$ originating from a left invariant Lagrangian, the reduced equations are $\dot \nu = {\rm ad}^*_{\chi(\nu)}\nu - i_{\chi(\nu)} \Sigma_\delta$ (here the 1-cocycle satisf\/ies $\sigma_\delta(g) = \delta(g)$).

\section{Routh reduction by stages}\label{section6}
In reduction by stages, we study the reduction of a $G$-invariant system (symplectic or Lagrangian) under the action of the full group $G$ and under the induced action w.r.t.\ a normal subgroup $K\lhd G$. We shall adopt as far as possible the notations used in~\cite{MarsdenHamRed}. A detailed construction of the following def\/initions is found in this reference.
\begin{definition}\mbox{} \label{def:notations}
\begin{enumerate}\itemsep=0pt
\item The Lie-algebra of $K$ is $\lak$ and $i$ denotes the injection $i: \lak\to \lag$ with dual $i^*: \lag^*\to \lak^*$.
\item The group $G$ acts on $\lak$ by restriction of the adjoint action. The induced action of $G$ on $\lak^*$ is denoted by the same symbol ${\rm Ad}^*:G\times\lak^*\to \lak^*$.
\item  $\mu$ denotes an element in $\lag^*$ and $\nu\in\lak^*$. Then $G_\mu$ is the isotropy subgroup of $\mu$ under the ${\rm Ad}^*$-action of $G$ on $\lag^*$; $G_\nu$ is the isotropy subgroup of $\nu$ under the ${\rm Ad}^*$-action of $G$ on $\lak^*$ obtained as the dual of the restricted $Ad$-action of $G$ on $\lak$; and $K_\nu$ is the isotropy of $\nu$ w.r.t.\ to standard coadjoint action of $K$ on $\lak^*$. These groups satisfy $G_\nu\cap K= K_\nu$ and~$K_\nu$ is normal in $G_\nu$.
\item $\lag_\nu$ and $\lak_\nu$ denote the Lie algebras of $G_\nu$ and $K_\nu$ respectively. $\bar G_\nu$ denotes the quotient group $G_\nu/K_\nu$ and its Lie algebra equals $\bar\lag_\nu= \lag_\nu/\lak_\nu$.
\item The projections onto the quotient groups are denoted by $r: G\to \bar G=G/K$ and $r_\nu:G_\nu\to\bar G_\nu$, and on the level of the Lie algebra: $r': \lag \to \bar \lag =\lag/\lak$ and $r'_\nu:\lag_\nu \to \bar\lag_\nu$.   The inclusion map $G_\nu \to G$ induces a map $k_\nu: \lag_\nu \to \lag$, with its dual $k_\nu^*: \lag^*\to \lag_\nu^*$.
\item $\rho$ denotes an element in $\bar\lag_\nu^*$.
\end{enumerate}
\end{definition}

In~\cite{MarsdenHamRed} symplectic reduction by stages is performed under the condition of a so-called `stages hypothesis'. An element $\mu\in\lag^*$ is said to satisfy the stages hypothesis if for any $\mu'\in\lag^*$ satisfying $\mu|_\lak=\mu'|_\lak=\nu$ and $\mu|_{\lag_\nu} = \mu'|_{\lag_\nu} =\bar\nu$, there exists an element $k\in K_\nu$ and $g\in (G_\nu)_{\bar\nu}$ such that ${\rm Ad}^*_{kg} \mu'= \mu$.
The stages hypothesis is a condition on a chosen momentum value and depends only on the symmetry group $G$. It was already clear in~\cite{MarsdenHamRed} that the hypothesis is automatically satisf\/ied if $G$ is a central extension or if $G$ is a semi-direct product group. In the recent contribution~\cite{mikityuk} it has been pointed out that the hypothesis is in fact always satisf\/ied, and that it can be taken out of the reduction by stages statements altogether. Taking advantage of this result, in this paper, we will not make further reference to the stages hypothesis.

\subsection{Symplectic reduction by stages}\label{section6.1}

\begin{figure}[htb]
\centering
\includegraphics{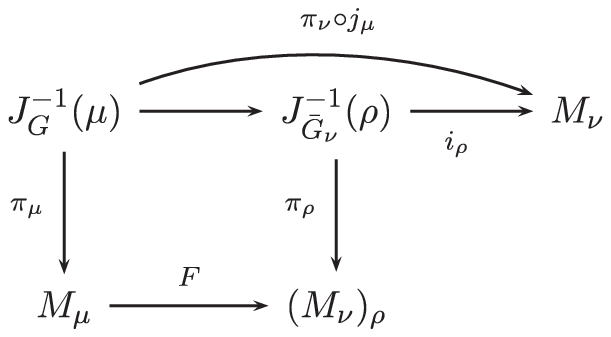}

\caption{Commuting diagram relating the dif\/ferent reduced symplectic manifolds.}\label{fig:diag5}\end{figure}

\begin{theorem}[Symplectic reduction by stages~\cite{MarsdenHamRed}]\label{thm:hamredstages}
Let $(M,\omega)$ be a symplectic manifold with a~canonical $G$-action $\Psi^M$ with an equivariant momentum map $J_G$.
\begin{enumerate}\itemsep=0pt
\item[$1.$] Fix a regular value $\mu\in\lag^*$ of the momentum map and perform symplectic reduction to obtain the symplectic manifold $(M_\mu,\omega_\mu)$.
\item[$2.$] The restriction of the action $\Psi^M$ to $K$ is canonical and the map $J_K=i^*\circ J_G:M\to \lak^*$ determines an equivariant momentum map for this induced action. Fix a regular value $\nu$ of $J_K$ and perform symplectic reduction to obtain the symplectic manifold $(M_\nu,\omega_\nu)$.

\item[$3.$] The level set $J^{-1}_K(\nu)$ is $G_\nu$-invariant.
\item[$4.$] The group $\bar G_\nu =G_\nu/K_\nu$ acts on $M_\nu$ by projecting the restricted action of $G_\nu$ on $J^{-1}_K(\nu)$. This induced action $\Psi^{M_\nu}$ is free, proper and canonical on $(M_\nu, \omega_\nu)$. Assume that $K_\nu$ is connected.
\item[$5.$] Fix an element $\bar \nu$ in $\lag_\nu^*$ such that the restriction of $\bar \nu|_{\lak_\nu}$ equals $\nu|_{\lak_\nu}$. There is a well-defined momentum map $J_{\bar G_\nu}: M_\nu \to \bar\lag^*_\nu$  for the induced action $\Psi^{M_\nu}$. This momentum map is determined from $J_G$ and $\bar \nu$ and has a non-equivariance cocycle:  $J_{\bar G_\nu}$ satisfies
\[(r'_\nu)^* \circ J_{\bar G_\nu} \circ \pi_\nu = k_\nu^*\circ J_G \circ i_\nu  -\bar \nu.\]
\item[$6.$] Let $\rho\in \bar\lag^*_\nu$ be a regular value for the momentum map $J_{\bar G_\nu}$ and let $(\bar G_{\nu})_\rho$ be the isotropy subgroup of $\rho$ w.r.t.\ the affine action of $\bar G_\nu$ on $\bar\lag_\nu^*$. Perform symplectic reduction to obtain the symplectic manifold $((M_\nu)_\rho, (\omega_\nu)_\rho)$ with $(M_\nu)_\rho = J^{-1}_{\bar G_\nu}(\rho)/(\bar G_\nu)_\rho$.
 \end{enumerate}
  If $\rho$ is chosen such that $(r'_\nu)^*\rho = \mu|_{\lag_\nu} - \bar \nu$, then there exists a symplectic diffeomorphism \[F: \ (M_\mu,\omega_\mu)\to ((M_\nu)_\rho, (\omega_\nu)_\rho).\]
\end{theorem}
For our purpose it is also important to understand the reduction of a $G$-invariant Hamiltonian~$h$ on~$M$. We assume that all conditions in Theorem~\ref{thm:hamredstages} are satisf\/ied. First note that, by def\/inition of the momentum maps, we have an inclusion $j_\mu$ of $J^{-1}_{G}(\mu)$ in $J^{-1}_K(\nu)$. Recall that we use $\pi_\nu$ for the projection $J^{-1}_K(\nu) \to M_\nu$. It was shown in~\cite{MarsdenHamRed} that the image of $\pi_\nu\circ j_\mu: J^{-1}_G(\mu) \to M_\nu$ is contained in $J^{-1}_{\bar G_\nu}(\rho)$. Moreover, this map is equivariant w.r.t.\ the action of $G_\mu$ on $J^{-1}_G(\mu)$ and $(\bar G_\nu)_\rho$ on $J^{-1}_{\bar G_\nu}(\rho)$ (this makes sense, since $G_\mu$ projects to a subset of~$(\bar G_\nu)_\rho$). The quotient of $\pi_\nu\circ j_\mu$ is the symplectic dif\/feomorphism $F$ mentioned in the previous theorem (see also Fig.~\ref{fig:diag5}).

Let $h$ be a $G$-invariant Hamiltonian on $M$ and let $h_\mu$ be the function on $M_\mu$ obtained from $\pi_\mu^*h_\mu = i^*_\mu H$. On the other hand we let $h_\nu$ be the function satisfying $\pi^*_\nu h_\nu = i^*_\nu h$. This function is $\bar G_\nu$-invariant: $h_\nu([m]_{K_\nu} [g]_{K_\nu} ) = h(i_\nu(mg)) = h(i_\nu(m) g)$, with $m\in J^{-1}_K(\nu)$ and $g\in G_\nu$ arbitrary. Note that $(\pi_\nu\circ j_\mu)^* h_\nu = i^*_\mu h$.

The Hamiltonian $h_\nu$ is a $\bar G_\nu$-invariant function on $(M_\nu, \omega_\nu)$. Applying the second symplectic reduction to this manifold, we obtain a new reduced Hamiltonian $(h_\nu)_\rho$ on $(M_\nu)_\rho$.

\begin{proposition}\label{prop:hamred} $F^* ((h_\nu)_\rho) = h_\mu$.\end{proposition}
\begin{proof}
We rely on the commuting diagram in Fig.~\ref{fig:diag5}: \[\pi_\mu^*(F^* ((h_\nu)_\rho)) = (\pi_\nu\circ j_\mu)^* h_\nu = i^*_\mu (h).\] This uniquely characterizes $F^* ((h_\nu)_\rho)$ as the function $h_\mu$.
\end{proof}

\subsection{Routh reduction by stages}\label{section6.2}
Routh reduction by stages is symplectic reduction by stages applied to the symplectic structure of the initial Lagrangian system. In this section we show that the symplectic structures and energy hamiltonians in the dif\/ferent stages can in fact be associated to specif\/ic magnetic Lagrangians systems, and eventually gives us Routh reduction by stages. The symplectic reduction by stages then provides us a dif\/feomorphism relating the solutions of the dif\/ferent Euler--Lagrange equations for the Lagrangian systems in the f\/inal stages.

We start with a hyperregular Lagrangian $L$, invariant under the action of a Lie group $G$. We assume that this Lagrangian satisf\/ies a regularity condition which is more stringent than mere $G$-regularity.
\begin{definition} The Lagrangian $L$ is said to be $G$-hyperregular if for any $v_q\in TQ$ and any subspace $\lak' <\lag$ with injection $i':\lak'\to \lag$,  the mapping ${i'}^*\circ J_L|_{v_q}\circ i':\lak' \to \lak'^*$ def\/ined by $\xi \mapsto {i'}^*\big(J_L(v_q+(i'(\xi))_Q(q))\big)$ is invertible.
\end{definition}
Lagrangians of mechanical type are $G$-hyperregular. Let $K$ be a normal subgroup of $G$, $\lak$ the Lie algebra of $K$ and $i:\lak\to \lag$ the canonical injection. Due to the hyperregularity the invariant Lagrangian~$L$ is both $G$- and $K$-hyperregular, and both $G$- and $K$-invariant. By def\/inition of~$J_L$, the map $i^*\circ J_L= i^*\circ (\psi^{TQ})^*\circ \F L$ is the momentum map for the $K$-action.

\begin{theorem}[Routh reduction by stages]\label{thm:routhstages}
Assume $(Q,L)$ is a hyperregular, $G$-hyperregular and $G$-invariant Lagrangian system. Let $K$ denote a normal subgroup of $G$.
\begin{enumerate}\itemsep=0pt
\item[$1.$]%\label{stages:1}
Let $\mu\in\lag^*$ be  a regular value of the momentum map $J_L$ and $\A^0$ a $G$-connection on $Q$. Let $(Q/G_\mu \to Q/G,  L_0, \B_0)$ be the magnetic Lagrangian system obtained by performing Routh reduction with respect to $G$.
\item[$2.$]%\label{stages:2}
Fix a regular value $\nu\in \lak^*$ of the momentum map $i^*\circ J_L$ for the $K$-action and a $K$-connection $\A^1$ on $Q$. Assume that $\A^1$ is $G$-equivariant w.r.t.\ the action of $G$ on $Q$ and $\lak^*$. Consider the magnetic Lagrangian system  $(Q/K_\nu \to Q/K,  L_{1},  \B_{1})$ obtained by performing Routh reduction with respect to $K$.

\item[$3.$]%\label{stages:4}
 $\bar G_\nu$ acts on $Q/K_\nu$ and $Q/K$ by projecting the induced action of $G_\nu$ on $Q$. These induced actions are free and proper.
\item[$4.$]%\label{stages:5}
Assume that $K_\nu$ is connected. Fix an element $\bar \nu\in \lag^*_\nu$ such that $\bar \nu |_{\lak_\nu} = \nu|_{\lak_\nu}$. Then the magnetic Lagrangian system  $(Q/K_\nu \to Q/K,  L_{1},  \B_{1})$ is $\bar G_\nu$-invariant, $\bar G_\nu$-regular and admits a $\B_1\bar\lag_\nu$-potential $\delta_1$ entirely determined by the choice of $\bar \nu$. The potential satisfies, for arbitrary $q\in Q$
\[
(r'_\nu)^*\big(\delta_1([q]_{K_\nu})\big)=  -(\psi^Q\circ k_\nu)^*\big(%Ê
\langle \nu , \A^{1}(q)\rangle\big) + \overline\nu .
\] Let $J_1$ denote the momentum map associated with $\delta_1$.
\item[$5.$]%\label{stages:6}
Fix a regular value $\rho\in \bar\lag_\nu^*$ for the momentum map $J_{1}$ and let $(\bar G_{\nu})_\rho$ be the isotropy subgroup of $\rho$ w.r.t.\ the affine action of $\bar G_\nu$ on $\bar\lag_\nu$. Fix a $\bar G_\nu$-connection on $Q/K$. Consider the magnetic Lagrangian system $((Q/K_\nu)/(\bar G_\nu)_\rho \to (Q/K)/\bar G_\nu), L_{2}, \B_{2})$ obtained by performing Routh reduction with respected to $\bar{G}_\nu$.
%\item\label{stages:7} Assume that $\mu$ satisfies the stages hypothesis.
\end{enumerate}
If $\rho$ is chosen such that $(r'_\nu)^*\rho = \mu|_{\lag_\nu} - \bar \nu$, then every solution $\gamma(t)\in Q/G_\mu$ to the Euler--Lagrange equations for $(Q/G_\mu\to Q/G, L_{0} , \B_{0})$ is mapped to a solution in $(Q/K_\nu)/(\bar G_\nu)_\rho$ to the Euler--Lagrange equations for $((Q/K_\nu)/(\bar G_\nu)_\rho \to (Q/K)/\bar G_\nu, L_{2}, \B_{2})$. Conversely, a solution in $(Q/K_\nu)/(\bar G_\nu)_\rho$ to  the Euler--Lagrange equations for $((Q/K_\nu)/(\bar G_\nu)_\rho \to (Q/K)/\bar G_\nu, L_{2}, \B_{2})$ is the projection of a solution in $Q/G_\mu$ to the Euler--Lagrange equations for $(Q/G_\mu\to Q/G, L_{0} , \B_{0})$.
\end{theorem}
\begin{proof}
 1 and 2 are obtained by applying Routh reduction. 3~follows from~\cite[p.~152]{MarsdenHamRed}: we know that the quotient groups $\bar G_\nu=G_\nu/K_\nu$ acts in a~free and proper way on the quotient space $Q/K_\nu$.  The group $\bar G_\nu$ is a subgroup of $\bar G$ and acts freely and properly and $Q/K$. We now show~4.

{\bf $\bar G_\nu$-Invariance of the Routh reduced system $(Q/K_\nu\to Q/K, L_1,\B_1)$.}

\begin{lemma} If the connection $\A^{1}$ is chosen such that it is equivariant w.r.t.\ the action of the full group $G$, i.e.\ if
\[
(\Psi^Q_g)^*\A^{1} = {\rm Ad}_{g^{-1}}\A^{1},
\]then the magnetic Lagrangian system $(Q/K_\nu\to Q/K, L_1,\B_1)$ is $\bar G_\nu$-invariant and $\delta_1$ is a $\B_1\bar\lag_\nu$-potential.
\end{lemma}
\begin{proof}
  We f\/irst show that $L_1$ is $\bar G_\nu$-invariant. For that purpose, we choose an arbitrary $\bar g\in \bar G_\nu$ and let $g\in G_\nu$ be a representative. Similar we choose a point $(v_{[q]_K},[q]_{K_\nu})\in T_{Q/K_\nu}(Q/K)$ such that it is the projection of $v_q\in (i\circ J_L)^{-1}(\nu)\subset TQ$. By def\/inition of the quotient action on $T_{Q/K_\nu}(Q/K)$, the action of $\bar g$ on an element $(v_{[q]_K},[q]_{K_\nu})$ equals the projection of $v_qg$. We now check the invariance of $L_1$ at an arbitrary point in $T_{Q/K_\nu}(Q/K)$:
  \begin{gather*}
    L_1\left(\Psi^{T_{Q/K_\nu}(Q/K)}_{\bar g}\big(v_{[q]_K},[q]_{K_\nu}\big)\right)   = L(\Psi^{TQ}_g(v_q) )- \langle \nu,\A^{1}(qg)(T\Psi^{Q}_g(v_q) )\rangle \\ \phantom{L_1\left(\Psi^{T_{Q/K_\nu}(Q/K)}_{\bar g}\big(v_{[q]_K},[q]_{K_\nu}\big)\right)}{}
    = L(v_q) - \langle {\rm Ad}_{g^{-1}}^*\nu , \A^{1}(q)(v_q) \rangle  = L_1\left(v_{[q]_K},[q]_{K_\nu}\right).
  \end{gather*}
Next, we check the $\bar G_\nu$-invariance of $\B_1$.  Recall that $\B_1$ is the projection to $Q/K_\nu$ of the 2-form $d\langle\nu,\A^{1}\rangle$ on $Q$. We f\/irst consider the equivariance of this 2-form under $G_\nu$. Let $g\in G_\nu$ be arbitrary, then
  \[
  (\Psi^{Q}_{g})^*(d\langle\nu,\A^{1}\rangle) =  d\langle {\rm Ad}^*_{g^{-1}}\nu ,\A^1\rangle = d\langle\nu,\A^{1}\rangle.
  \]We thus obtain $G_\nu$-invariance for $d\langle\nu,\A^{1}\rangle$, and we may conclude that $(\Psi^{Q/K_\nu}_{\bar g})^* \B_1 = \B_1$ holds on $Q/K_\nu$.

  The third and f\/inal step is the def\/inition of the $\B_1\bar\lag_\nu$-potential. We consider an element $\bar \xi = [\xi]_{\lak_\nu} \in \bar\lag_\nu=\lag_\nu/\lak_\nu$ and let $\xi \in \lag_\nu$ be a representative. Then, by def\/inition of $\B_1$, the 1-form $i_{\bar \xi_{Q/K_\nu}}\B_1$ is the projection to $Q/K_\nu$ of the 1-form $i_{\xi_Q}d(\langle\nu,\A^{1}\rangle)$ on $Q$ (i.e.\ $\xi_Q$ projects to $\bar \xi_{Q/K_\nu}$).
  Again we concentrate on the 1-form on $Q$: \[
  i_{\xi_Q}d\langle\nu,\A^{1}\rangle = {\cal L}_{\xi_Q}(\langle\nu,\A^{1}\rangle) - d\big(i_{\xi_Q}\langle\nu,\A^{1}\rangle\big).\] Since $\langle\nu,\A^{1}\rangle$ is $G_\nu$-invariant, we conclude that $i_{\xi_Q}d\langle\nu,\A^{1}\rangle = - d\big(i_{\xi_Q}\langle\nu,\A^{1}\rangle\big)$. The exact 1-from on the right gives a strong hint of the structure of the $\bar\lag_\nu$-potential. Assume now that we f\/ixed an element $\overline\nu \in \lag_\nu^*$ such that $\overline\nu|_{\lak_\nu}=\nu|_{\lak_\nu}$.

  The function $\delta$  on $Q$,  def\/ined by
  \[-\delta_\xi(q) = \langle\nu,\A^{1}(q)(\xi_Q(q))\rangle - \langle \overline\nu,\xi\rangle \] is our candidate for the $\B_1\bar\lag_\nu$-potential. This statement makes sense provided that $\delta_\xi$ projects to a function on $Q/K_\nu$ and that it only depends on the equivalence class $\bar \xi=\xi+\lak_\nu$ of $\xi\in\lag_\nu$. The latter is a straightforward consequence of the fact that $\A^{1}$ is a principal $K$-connection. The $K_\nu$-invariance is more involved, and we rely on a result in~\cite{MarsdenHamRed}. For any $k$ in $K_\nu$, we have
  \begin{gather*}
  - \delta_{\xi}(qk)=  \langle \nu, \A^{1}(qk)(\xi_Q(qk))\rangle - \langle \overline\nu,\xi\rangle = \langle \nu, \A^{1}(q)(({\rm Ad}_{k}\xi)_Q(q))\rangle - \langle \overline\nu,\xi\rangle.
  \end{gather*}
  Therefore $\delta_\xi$ is constant on the orbits of $K_\nu$ in $Q$ if
  $\langle \nu, \A^{1}(q)((\xi-{\rm Ad}_{k}\xi)_Q(q))\rangle$ vanishes for all~$k$. To show this we introduce a function $f$ on $K_\nu$ given by $f(k)=\langle \nu, \A^{1}(q)((\xi-{\rm Ad}_{k}\xi)_Q(q))\rangle$ and we use similar arguments as in~\cite[p.~156]{MarsdenHamRed}. If we can show that $f(e)=0$,  $df|_e=0$ and $f(k_1k_2)=f(k_1)+f(k_2)$ for arbitrary $k_{1,2}\in K_\nu$, we may conclude that  $f=0$ (since $K_\nu$ is assumed connected).

The f\/irst condition $f(e)=0$ is trivial. To check the second condition: let $\kappa \in \lak_\nu$ be arbitrary, then
  \[
  df|_e(\kappa) = \langle \nu, \A^{1}(q)(-{\rm ad}_{\kappa}\xi)_Q(q))\rangle = -\langle \nu, {\rm ad}_\kappa\xi\rangle  = -\langle {\rm ad}^*_\kappa,\xi\rangle=0.
  \]
Above, we have used the fact that $K_\nu$ is normal in $G_\nu$ and that, as a consequence, the Lie bracket $[\kappa,\xi]$ is in $\lak_\nu$. Therefore the contraction of the corresponding fundamental vector f\/ield with $\A^{1}$ is precisely $[\kappa,\xi]$. Next, we check the third condition and compute $f(k_1k_2)$. Given the identity
  \[
  \xi-{\rm Ad}_{k_1}{\rm Ad}_{k_2} \xi =\xi -{\rm Ad}_{k_1}\xi+{\rm Ad}_{k_1}(\xi-{\rm Ad}_{k_2}\xi)
  \]
  and the fact that $k_1\in K_\nu$,
  \begin{gather*}
    f(k_1k_2) = \langle \nu, \A^{1}(q)((\xi-{\rm Ad}_{k_1k_2}\xi)_Q(q))\rangle\\
    \phantom{f(k_1k_2)}{} = \langle \nu, \A^{1}(q)((\xi-{\rm Ad}_{k_1}\xi)_Q(q)) \rangle + \langle \nu, \A^{1}(q)({\rm Ad}_{k_1}(\xi-{\rm Ad}_{k_2}\xi)_Q(q))\rangle\\
    \phantom{f(k_1k_2)}{} = f(k_1) +  \langle {\rm Ad}_{k_1}^*\nu, \A^{1}(q)((\xi-{\rm Ad}_{k_2}\xi)_Q(q))\rangle = f(k_1)+f(k_2).
  \end{gather*}
This completes the proof: the $\lag_\nu^*$-valued function $\delta$ is shown to be projectable to a $\bar\lag_\nu^*$-valued function on $Q/K_\nu$.  This is the sought-after potential $\delta_1$: for arbitrary $q\in Q$, we have \[
(r'_\nu)^*\big(\delta_1([q]_{K_\nu})\big)=  -(\psi^Q\circ k_\nu)^*\big(%Ê
\langle \nu , \A^{1}(q)\rangle\big) + \overline\nu . \tag*{\qed}
\]\renewcommand{\qed}{}
\end{proof}

{\bf Symplectic structure of $(Q/K_\nu\to Q/K, L_1,\B_1)$ and symplectic reduction by stages.}
\begin{lemma} Apply symplectic reduction by stages to the symplectic structure associated to the $G$-invariant Lagrangian system $(Q,L)$. Identify the symplectically reduced manifold $M_\nu$ with the symplectic structure on $T_{Q/K_\nu}(Q/K)$ induced by the magnetic Lagrangian system $(Q/K_\nu\to Q/K, L_1,\B_1)$. Then:
\begin{enumerate}\itemsep=0pt
\item[$1.$] The action $\Psi^{T_{Q/K_\nu}(Q/K)}$ of $\bar G_\nu$ on $T_{Q/K_\nu}(Q/K)$ is precisely the induced action on the first reduced space $M_\nu$ in symplectic reduction by stages.
\item[$2.$]  For a chosen $\bar \nu\in \lag_\nu^*$, the momentum map $J_{1}: T_{Q/K_{\nu}}(Q/K)\to \bar\lag_\nu^*$ associated with the magnetic Lagrangian system $(Q/K_\nu\to Q/K, L_1,\B_1)$ corresponds to the induced momentum map $J_{\bar G_\nu}$ from symplectic reduction by stages.
\end{enumerate}
\end{lemma}
\begin{proof}
1. The momentum map for the $K$ action is precisely $J_K:=i^*_\lak \circ J_L$, with $J_L:TQ\to \lag^*$. By def\/inition, the induced action of $\bar G_\nu$  on $J_K^{-1}(\nu)/K_\nu$ is obtained by projecting the action of~$G_\nu$ on~$J_K^{-1}(\nu)$. If we take into account that we realize the quotient manifold $J_K^{-1}(\nu)/K_\nu$ as $T_{Q/K_\nu}(Q/K)$, the induced action on $T_{Q/K_\nu}(Q/K)$ is obtained by projection of the action of $G_\nu$ on $TQ$ under the projection $TQ \to  T_{Q/K_\nu}(Q/K)$. This is precisely the action we have introduced above.

2. The {\em induced momentum map} ${J}_{\bar G_\nu}$ is def\/ined in the following way (we consider it directly as a function on $T_{Q/K_{\nu}}(Q/K)$ instead of on $M_\nu$):
\begin{gather}\label{eq:mommap}
\left\langle{J}_{\bar G_\nu}\big(v_{[q]_K}, [q]_{K_\nu}\big), \bar\xi\right\rangle = \langle J_L(v_q),k_\nu(\xi)\rangle - \langle \bar\nu,\xi\rangle = \langle \F L(v_q),\xi_Q\rangle - \langle \bar\nu,\xi\rangle,
\end{gather}
where $\xi \in \lag_\nu$ is arbitrary and projects to $\bar \xi \in \bar \lag_\nu$, $v_q$ projects to $(v_{[q]_K}, [q]_{K_\nu})$.  By def\/inition of the momentum map of the magnetic Lagrangian system $(Q/K_\nu\to Q/K, L_1,\B_1)$, we have
\begin{gather}\label{eq:mommap2}
\left\langle{J}_{1}\big(v_{[q]_K}, [q]_{K_\nu}\big), \bar\xi\right\rangle = \left\langle \F L_1\big(v_{[q]_K}, [q]_{K_\nu}\big),\big( (\bar \xi)_{Q/K},[p]_{K_\nu}\big)\right\rangle -  (\delta_1)_{\bar \xi}([p]_{K_\nu}).
\end{gather}
We now show that the right-hand side of~\eqref{eq:mommap2} equals the right-hand side of~\eqref{eq:mommap}. We therefore use the def\/inition of $\F L_1$ and $\delta_1$ as being the projection of maps upstairs:
\begin{gather*}
   \left\langle \F L_1\big(v_{[q]_K}, [q]_{K_\nu}\big),\big( (\bar \xi)_{Q/K},[q]_{K_\nu}\big)\right\rangle= \langle \F L(v_q),\xi_Q(q)\rangle - \langle \nu,\A^1(q)(\xi_Q(q))\rangle , \\
   (\delta_1)_{\bar \xi}([q]_{K_\nu})= \delta_\xi(p)= - \langle \nu,\A^1(q)(\xi_Q(q))\rangle + \langle \overline\nu,\xi\rangle. \tag*{\qed}
\end{gather*}\renewcommand{\qed}{}
\end{proof}

Finally, before we can reapply Routh reduction for the second stage, we need to check that $L_1$ is $\bar G_\nu$-regular.
\begin{lemma} The magnetic Lagrangian system $(Q/K_\nu\to Q/K, L_1,\B_1)$ is $\bar G_\nu$-regular.
\end{lemma}
\begin{proof}
  We have to show that, for any $(v_{[q]_K},[q]_{K_\nu})$ the map  \[\bar \lag_\nu \to \bar\lag_\nu^*; \qquad \bar \xi \mapsto {J}_{1}(v_{[q]_K}+\bar \xi_{Q/K}([q]_K),[q]_{K_\nu})\] is invertible. Let $v_q$ determine a tangent vector in $J^{-1}_K(\nu)$ representing $(v_{[q]_K},[q]_{K_\nu})$. Let $\eta$ denote an arbitrary element in $\bar \lag_\nu^*$. Due to the assumed $G$-hyperregularity, there is a unique $\xi\in\lag_\nu$ such that $k_\nu^*(J_L(v_q+ \xi_Q(q)))=(r'_\nu)^*\eta + \overline\nu$. The projection $\bar\xi=r'_\nu(\xi)$ of $\xi$ def\/ines the inverse element for $\eta$, since it is such that \[k^*_{\nu}(J_L(v_q+ \xi_Q(q))) - \bar\nu = (r'_\nu)^*{J}_{1} (v_{[q]_K}+\bar \xi_{Q/K})=(r'_\nu)^*\eta. \tag*{\qed}\] \renewcommand{\qed}{}
\end{proof}

{\bf Symplectic and Routh reduction by stages.}
Summarizing the above lemmas, we conclude that the magnetic Lagrangian system $(Q/K_\nu\to Q/K, L_1,\B_1)$ is amenable to Routh reduction and that the symplectic structure and momentum map associated to this Lagrangian system correspond to the symplectic structure and momentum map encountered in symplectic reduction by stages. If $\rho$ is chosen such that the compatibility relation $(r'_\nu)^*\rho = \mu|_{\lag_\nu} - \bar \nu$ holds, then from symplectic reduction by stages we have that the symplectic structures associated to $(Q/G_\mu\to Q/G, L_{0} , \B_{0})$ and $((Q/K_\nu)/(\bar G_\nu)_\rho \to (Q/K)/\bar G_\nu, L_{2}, \B_{2})$ are symplectically dif\/feomorphic by means of the symplectic dif\/feomorphism $F$ introduced earlier.  From  Proposition~\ref{prop:hamred} it follows that $F^*E_{L_2} = E_{L_0}$ and therefore the corresponding Hamiltonian vector f\/ields are $F$-related.  We def\/ine a map $\tau: Q/G_\mu \to (Q/K_\nu)/{(\bar G_\nu)_\rho}$ as $\tau([q]_{G_\mu}) = \big[[q]_{K_\nu}\big]_{(\bar G_\nu)_\rho}$. The map is well-def\/ined since $G_\mu$ is a subgroup of $G_\nu$ and since $r_\nu(G_\mu) \subset (\bar G_\nu)_\rho$.
\begin{figure}[htb]
\centering
\includegraphics{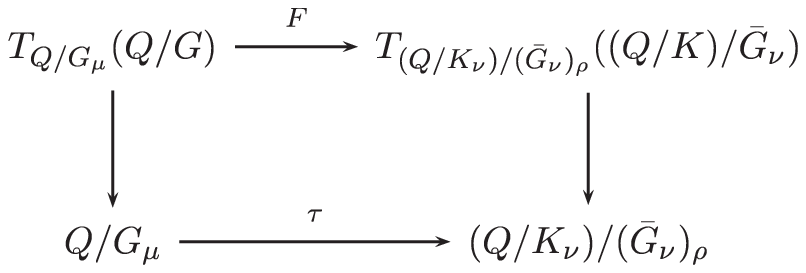}

\caption{Fibration of the symplectic dif\/feomorphism $F$.}\label{fig:diag4}
\end{figure}

\begin{lemma} The symplectic diffeomorphism $F$ is fibred over $\tau$.\end{lemma}
\begin{proof} We recall the def\/inition of the map $F$: f\/ix an element $(v_{[q]_G}, [q]_{G_\mu})$ and let $v_q\in J^{-1}_L(\mu)$ be a representative.  The point $F (v_{[q]_G}, [q]_{G_\mu})$ is obtained by taking the consecutive quotients of~$v_q$. In particular, the component of the f\/inal quotient in the conf\/iguration space $(Q/K_\nu)/(\bar G_\nu)$ of the magnetic Lagrangian system, is precisely the image $\tau$. \end{proof}

Since $F$ is a dif\/feomorphism, $\tau$ is onto. The Hamiltonian vector f\/ield on $T_{Q/G_\mu}(Q/G)$ and $T_{(Q/K_\nu)/(\bar G_\nu)_\rho}((Q/K)/\bar G_\nu)$ are $F$-related. Their integral curves project onto solutions of the Euler--Lagrange equations. This concludes the proof of Theorem~\ref{thm:routhstages}.
\end{proof}

\section{Examples}\label{section7}
\subsection{Elroy's Beanie}\label{section7.1}
This system appears in e.g.~\cite{marsdenphases}. It consists of two planar rigid bodies that
are connected in their center of mass. The system moves in the plane
and it is subject to some conservative force with potential $V$. The
conf\/iguration space is $SE(2)\times S^1$, with coordinates $(x,y,\theta,\psi)$. Here $(x,y)$ is the position of the center of
mass, $\theta$ is the rotation of the f\/irst rigid body, and $\psi$
the relative rotation of the second body w.r.t.\ the f\/irst. The
kinetic energy of the system is $SE(2)$-invariant and we will
suppose that the potential is invariant as well. This means in fact that only the
relative position of the two bodies matters for the dynamics of the
system. The Lagrangian is of the form
\[
L=\frac12 m ({\dot x}^2 + {\dot y}^2) + \frac12 I_1 {\dot\theta}^2 +
\frac12 I_2 (\dot\theta+\dot\psi)^2 -V(\psi).
\]
The Euler--Lagrange equations of the system are, in normal form,
\[
\ddot x =0, \quad \ddot y=0, \qquad \ddot\theta = \frac{1}{I_1} V',
\qquad \ddot\psi = - \frac{I_1+I_2}{I_1I_2} V'.
\]

{\bf The symmetry group and the principal connection.}
An element of $SE(2)$, the special Euclidean group, can be represented
by a matrix of the form
\begin{gather} \label{euclmatrix}
\left( \begin{array}{ccc} \cos\theta & -\sin\theta & x \\
\sin\theta &\cos\theta & y \\ 0&0&1 \end{array}\right).
\end{gather}
The identity of the group is $(x=0$, $y=0$, $\theta=0)$ and the
multiplication is given by
\[
(x_1,y_1,\theta_1) * (x_2,y_2,\theta_2) =
(x_2\cos\theta_1-y_2\sin\theta_1 + x_1,
x_2\sin\theta_1+y_2\cos\theta_1 + y_1, \theta_1+\theta_2).
\]

The matrices
\[
e_1=  \begin{pmatrix} 0 & 0 & 1 \\
0 &0& 0 \\ 0&0&0\end{pmatrix}, \qquad
e_2=  \begin{pmatrix} 0 & 0 & 0 \\
0 &0& 1 \\ 0&0&0\end{pmatrix}, \qquad
e_3=  \begin{pmatrix} 0 & -1 & 0 \\
1 &0& 0 \\ 0&0&0\end{pmatrix},
\]
form a basis for the Lie algebra, for which $[e_1,e_2]=0$,
$[e_1,e_3]=e_2$ and $[e_2,e_3]=-e_1$.  A~corresponding basis of fundamental vector f\/ields on $Q$ is
\[
{\tilde e}_1 = \fpd{}{x}, \qquad {\tilde e}_2 = \fpd{}{y}, \qquad
{\tilde e}_3 = - y\fpd{}{x}+x\fpd{}{y} +\fpd{}{\theta},
\]
and a basis of invariant vector f\/ields is
\[
{\hat e}_1 = \cos\theta\fpd{}{x}+\sin\theta\fpd{}{y}, \qquad {\hat
e}_2 = -\sin\theta\fpd{}{x}+\cos\theta \fpd{}{y} , \qquad {\hat e}_3
= \fpd{}{\theta}.
\]
One can easily verify that the Lagrangian is invariant under the $SE(2)$-action. There is a trivial
principal connection on $P=Q=SE(2)\times S^1\to Q/SE(2)=S^1$, which locally takes the form
\[
(dx+yd\theta) e_1 + (dy -x d\theta) e_2 + d\theta e_3.
\]
The momentum map $J_L$ is given by:
\[
J_L= m\dot x e^1 + m \dot ye^2 +\big(m (x\dot y - y \dot x) +I_1\dot \theta + I_2(\dot \theta+\dot \psi)\big) e^3.
\]

In what follows we perform two Routh reductions on  the Lagrangian. The f\/irst reduction is done w.r.t.\ the full symmetry group $SE(2)$, and the second reduction w.r.t.\ the Abelian normal subgroup $\R^2$.

{\bf Full reduction.} Let $\mu=\mu_1e^1+\mu_2e^2+\mu_3e^3$ be a generic element in $\lag^*$. An element $\xi=\xi^1e_1+\xi^2e_2+\xi^3e_3$ of the isotropy algebra
$\lag_\mu$ satisf\/ies
\[
\xi^3\mu_2 =0, \qquad \xi^3\mu_1 =0, \qquad
\xi^1\mu_2-\xi^2\mu_1 =0.
\]
So if we suppose that $\mu_1$ and $\mu_2$ do not both vanish~-- we will set $\mu_1=1$  from now on~-- then a~typical element of $\lag_\mu$ is $\xi= \xi^1(e_1+\mu_2 e_2)$. Since $\lag_\mu$ is
1-dimensional, $G_\mu$ is of course Abelian. A~convenient way to describe the manifold $P/G_\mu=SE(2)/G_\mu$ locally is by considering a coordinate transformation $(x',y',\theta')$ in the group coordinates such that the vector f\/ield associated to an element in $\lag_\mu$ becomes a coordinate vector f\/ield: in the new coordinates, we should have $\partial_{x'}=\partial_{x}+\mu_2\partial_{y}$.
 This is obtained by the following transformation
\[
x'=x, \qquad y'= y-\mu_2 x, \qquad \theta'=\theta.
\]
Then clearly $(y',\theta)$ is a coordinate chart on $SE(2)/G_\mu$. And simultaneously, we have that $(y',\theta,\psi)$ is a coordinate chart on the reduced conf\/iguration manifold $P/G_\mu = (SE(2)\times S^1)/G_\mu$, and the f\/ibration $P/G_\mu \to Q/G=S^1$ is locally represented by $(y',\theta,\psi)\mapsto (\psi)$. For a more systematic treatment on appropriate coordinate changes, we refer to~\cite{mestcram}.

We now compute the Routhian $L_0$ and the 2-form $\B_0$. Following~\cite{pars}, a convenient way to compute the (unreduced) Routhian for mechanical Lagrangians is by using $2( L_0 +V)=\big(-p_x\dot x -p_y\dot y - \dot \theta p_\theta + \dot \psi p_\psi\big)_{J^{-1}_L(\mu)}$, where $p_i$ is the momentum in the $i$th coordinate. We have:
\begin{gather*}
 2( L_0 +V)  = \left(-m ({\dot x}^2 + {\dot y}^2) -I_1 {\dot\theta}^2 -
 I_2 (\dot\theta+\dot\psi)\dot\theta  +I_2(\dot \theta+\dot\psi)\dot\psi\right)_{J^{-1}_L(\mu)}  \\
 \phantom{2( L_0 +V)}{}
 =  \left(-m ({\dot x}^2 + {\dot y}^2) -(I_1+I_2) {\dot\theta}^2 +I_2\dot\psi^2\right)_{J^{-1}_L(\mu)} \\
 \phantom{2( L_0 +V)}{}= - \frac{1}{m}(1+ \mu_2^2) -\frac{\left(\mu_3 -(x\mu_2-y) -I_2\dot\psi\right)^2}{I_1+I_2} +
I_2\dot\psi^2\\
\phantom{2( L_0 +V)}{}
=  \frac{I_1I_2}{I_1+I_2}\dot\psi^2 +2I_2\frac{\mu_3-(x\mu_2-y)}{I_1+I_2}\dot\psi - \frac{\left(\mu_3 -(x\mu_2-y)\right)^2}{I_1+I_2}.
\end{gather*}
In the last step we have left out some constant terms. The reduced Lagrangian is then obtained by taking the quotient w.r.t.\ the action of $G_\mu$.  This is done by applying the coordinate transformation introduced above. We get:
 \[
 L_0 =\frac12 \frac{I_1I_2}{I_1+I_2}\dot\psi^2 +I_2\frac{\mu_3+y'}{I_1+I_2}\dot\psi - \left(V(\psi) +\frac12 \frac{\left(\mu_3 +y'\right)^2}{I_1+I_2}\right),
\]
which is clearly independent of $x'$.
The 2-form $\B_0$ is obtained by reducing the 2-form
\[ d\left((dx+yd\theta) + \mu_2(dy-xd\theta) +\mu_3d\theta\right)= d(y-\mu_2 x)\wedge d\theta.\]
Using the coordinate change  we get $\B_0=dy'\wedge d\theta$.

In this example the Routhian $L_0$ depends on the velocity corresponding to the coordinate~$\psi$ on~$S^1$, but is independent of the velocities corresponding to the two remaining coordinates~$(y',\theta)$. With the above, the reduced Euler--Lagrange equations take the form
\begin{gather*}
{\dot y}'  =  0,\\
{\dot\theta}   =  \frac{1}{I_1+I_2} (y'+\mu_3-I_2\dot\psi),  \\
{\ddot \psi} =  - \frac{I_1+I_2}{I_1I_2} V' - \frac{1}{I_1} \dot y'.
\end{gather*}
Note that the second order equation in $\psi$ decouples from the f\/irst order equations, and that these two f\/irst order equations are the momentum equations rewritten in normal form.

{\bf Abelian reduction.} We now perform f\/irst Routh reduction w.r.t.\ the Abelian symmetry group $\R^2$ of translations in the $x$ and $y$ direction. Let us denote the symmetry group by $K=\R^2$ and study the quotient spaces. We will use the same notations as before: the Lie algebra elements $e_1$,  $e_2$ denote a basis for the subalgebra~$\lak$ of~$K$ in~$\lag$. The momentum map for this action is now $J_K=i^* \circ J_L =m\dot xe^1 + m\dot y e^2$. We choose $\nu$ to be the projection of the momentum~$\mu$ we had used in the full reduction: let $\nu = e^1 + \mu_2 e^2\in \lag^*$. Since $K$ is Abelian, $K_\nu=K$  and the quotient space is $SE(2)\times S^1/\R^2=S^1\times S^1$. If we choose $\A^1=dx e_1 +dy e_2 $ to be the trivial connection, we simply get $\B_1=0$. The Routhian $L_1$ can now be obtained from
\begin{gather*}
 2( L_1 +V)  = \left( -p_x\dot x- p_y \dot y + p_\theta \dot \theta +p_\psi \dot \psi\right)_{J^{-1}_L(\nu)} \\
 \phantom{2( L_1 +V)}{}
 = \left(-m\dot x^ 2- m\dot y^2 + I_1 {\dot\theta}^2 + I_2 (\dot\theta+\dot\psi)^2\right)_{J^{-1}_L(\nu)}
 = I_1 {\dot\theta}^2 +  I_2 (\dot\theta+\dot\psi)^2,
\end{gather*}
where we ignored again some constant terms. The Routh reduced system is now a standard Lagrangian system on $S^1\times S^1$ with Lagrangian $L_1 =  \frac12I_1 {\dot\theta}^2 +  \frac12I_2 (\dot\theta+\dot\psi)^2-V(\psi)$ (see the paragraphs on Abelian Routh reduction). Its equations of motion are
\begin{gather*}
{\ddot \theta}  =  \frac{1}{I_1} V', \qquad {\ddot \psi} =  -
\frac{I_1+I_2}{I_1I_2} V'.
\end{gather*}

For this example there is actually no second stage: the group ${\bar G}_\nu=G_\nu/K_\nu$ is the trivial one~$\{e\}$, and the vector space ${\bar \lag}_\nu=\lag_\nu/\lak_\nu$ is only the zero
vector. So, there is no second momentum map to take into account,
and there is no further symmetry to quotient out.

In the reduction by stages process we have not made use of $\mu_3$. We now show that the two ways of reducing the system are equivalent.

{\bf Equivalence between direct reduction and reduction by stages.}
Let us compute the dif\/feomorphism $F$ for this example. Here, it is a map $J^{-1}_L(\mu)/G_\mu \to J^{-1}_K(\nu)/K$ that is obtained by projection of the inclusion map $J^{-1}_L(\mu)\to  J^{-1}_K(\nu)$. The latter equals, in coordinates
\[
(x,y,\theta,\psi,\dot \psi) \mapsto \left(x,y,\theta,\psi,\dot \theta = \frac{1}{I_1+I_2}(\mu_3-(x\mu_2 - y)-I_2\dot\psi), \dot \psi\right).
\]
The map is reducible, and after taking the quotient it becomes
\begin{gather*}
F: \ J^{-1}_L(\mu)/G_\mu \to J^{-1}_K(\nu)/K;\qquad (y',\theta,\psi,\dot \psi) \mapsto \left(\theta,\psi,\dot \theta =  \frac{1}{I_1+I_2}(\mu_3+y'-I_2\dot\psi), \dot \psi\right).
\end{gather*}

This dif\/feomorphism maps the $G_\mu$-reduced system on the
$K_\nu$-reduced system, as is obvious from the respective equations of motion.

\subsection{Rigid bodies on the Heisenberg group}\label{section7.2}

As a second example of Routh reduction by stages, we discuss the dynamics of a rigid body immersed in a potential f\/low with circulation~\cite{VaKaMa2009b}.   We assume that the body is circular, and in this case the equations of motion are given by
\begin{gather} \label{circeqns}
	\frac{d}{dt}
		\begin{bmatrix}
			p_x \\ p_y
		\end{bmatrix} = \Gamma
			\begin{bmatrix}
				- v_y \\
				v_x
			\end{bmatrix},
	\qquad \text{where}\quad
	\begin{bmatrix}
			p_x \\ p_y
		\end{bmatrix} = \mathbb{M}
	\begin{bmatrix}
			v_x \\ v_y
		\end{bmatrix}.
\end{gather}
Here $\Gamma$ represents the circulation and $\mathbb{M}$ is a (non-diagonal) mass matrix, which incorporates the inertia and added masses of the body.  The right-hand side  of the equations of motion represents the so-called \emph{Kutta--Joukowski lift force}, a gyroscopic force due to circulation \cite{lamb, MiTh1968}.

While this system is extremely easy to integrate, it nevertheless exhibits all the interesting geometric characteristics of more complicated examples.  As we show below, the conf\/iguration space for this system is the Heisenberg group, arguably the simplest non-trivial central extension group, and the procedure of reduction by stages demonstrated here can be applied equally well to more complicated central extensions, such as the oscillator group (describing the dynamics of rigid bodies of arbitrary cross section in circulatory f\/low) and the Bott--Virasoro group describing the KdV equation.

In this context, the Heisenberg group $H$ is the Euclidian space $\mathbb{R}^3$, equipped with the multiplication
\[
	(x, y, s) \cdot (x', y', s') = \left(x + x', y + y', s + s' + \frac{1}{2} (xy' - yx')\right),
\]
using the def\/inition of $H$ given in \cite{Montboek}.  The coordinates $(x, y)$ describe the center of the circular disc, while the coordinate $s$ corresponds to the \emph{flux} of the f\/luid around the body.  The conjugate momentum corresponding to $s$ will be the circulation of the f\/luid.

The Heisenberg group can alternatively be described as the central extension of $\mathbb{R}^2$ constructed by means of the cocycle $B((x, y), (x', y')) = \frac{1}{2} (xy' - yx')$.  We now introduce the inf\/initesimal cocycle $C: \mathbb{R}^2 \times \mathbb{R}^2 \to \mathbb{R}$, given by
\[
	C((v_x, v_y), (v_x', v_y')) = v_x v_y' - v_y v_x'.
\]
For the relation between $B$ and $C$, see for instance \cite{MarsdenHamRed}. The Lie algebra $\mathfrak{h}$ of $H$ can then be identif\/ied with $\mathbb{R}^3$ with the bracket
\begin{gather*}
	[(v_x, v_y, v), (v_x', v_y', v')]   = (0, 0, C((v_x, v_y), (v_x', v_y')))
		  = (0, 0, v_x v_y' - v_y v_x').
\end{gather*}
The dual Lie algebra $\mathfrak{h}^\ast$ can again be identif\/ied with $\mathbb{R}^3$, with coordinates $(p_x, p_y, p)$ and duality pairing $\left< (p_x, p_y, p), (v_x, v_y, v) \right> = p_x v_x + p_y v_y + pv$.

We now induce the following quadratic Lagrangian $\ell$ on $\mathfrak{h}$:
\[
	\ell(v_x, v_y, v) = \frac{1}{2} \left[v_x, v_y\right] \mathbb{M} \begin{bmatrix} v_x \\ v_y \end{bmatrix} + \frac{v^2}{2},
\]
and we def\/ine $L$ on $H$ by left extension: $L(g, \dot{g}) = \ell(g^{-1} \dot{g})$, or explicitly
\begin{gather} \label{circlagr}
	L(x, y, s; \dot{x}, \dot{y}, \dot{s}) = \frac{1}{2} \left[\dot{x}, \dot{y} \right] \mathbb{M} \begin{bmatrix} \dot{x} \\ \dot{y} \end{bmatrix}
		+ \frac{1}{2} \left( \dot{s} - \frac{1}{2} (x \dot{y} - y \dot{x}) \right)^2.
\end{gather}

The Euler--Poincar\'e equations obtained from $\ell$ are given by
\[
	\frac{d}{dt} \begin{bmatrix}
		p_x \\ p_y \\ p
	\end{bmatrix} =
		p
	\begin{bmatrix}
		-v_y \\ v_x \\ 0
	\end{bmatrix}.
\]
After setting $p = \Gamma$, we  obtain the equations \eqref{circeqns}.  These equations also coincide with the Euler--Lagrange equations obtained from the Lagrangian \eqref{circlagr}.

{\bf Reduction with respect to $\mathbb{R}$.}  The center of $H$ is the normal subgroup isomorphic to~$\mathbb{R}$ which consists of all elements of the form $(0, 0, s)$, where $s \in \mathbb{R}$.   We f\/irst perform Routh reduction with respect to the left action of this subgroup on $H$.   On the principal bundle $H \to H/\mathbb{R} \cong \mathbb{R}^2$ we consider the connection one-form given at the identity by $\mathcal{A}(e)(v_x, v_y, v) = v$, and extended to the whole of $H$ by left translation.  Explicitly, we have
\[
	\mathcal{A}(x, y, s)
	= ds - \frac{1}{2} (x dy - ydx).
\]
Since the structure group $\mathbb{R}$ is Abelian, the curvature of $\mathcal{A}$ is given by $\mathcal{B} = d \mathcal{A} = - dx \wedge dy$.

Similarly, the momentum map $J_L: TH \to \mathbb{R}$ for the $\mathbb{R}$-action on the tangent bundle $TH$  is given by $J_L(x, y, z; \dot{x}, \dot{y}, \dot{z}) = \dot{s} - (x \dot{y} - y \dot{x})/2$, so that $J_L^{-1}$ consists of all points $(x, y, s; \dot{x}, \dot{y}, \dot{s})$ with $\dot{s} = \Gamma + (x \dot{y} - y \dot{x})/2$.  The isotropy subgroup $\mathbb{R}_\Gamma = \mathbb{R}$ acts on this level set by translations in the $s$-direction, so that the reduced velocity space is given by
\[
	J^{-1}(\Gamma) / \mathbb{R}_\Gamma = T \mathbb{R}^2.
\]
The symplectic form on the reduced space can easily be computed, and is explicitly given by
\[
	 A d\dot{x} \wedge dx + B(d\dot{x} \wedge dy + d\dot{y} \wedge dx) + C d\dot{y} \wedge dy - \Gamma dx \wedge dy,
\]
where $A$, $B$, $C$ are the entries of the mass matrix $\mathbb{M}$.  The last term of the symplectic form, $-\Gamma dx \wedge dy$, is the curvature term of the connection, paired with $\Gamma \in \mathbb{R}$.  Finally, a quick computation shows that the reduced Lagrangian is just the kinetic energy Lagrangian on $T \mathbb{R}^2$:
\begin{gather} \label{redlagr}
	L_1(x, y; \dot{x}, \dot{y}) = \frac{1}{2} \left[\dot{x}, \dot{y} \right] \mathbb{M} \begin{bmatrix} \dot{x} \\ \dot{y} \end{bmatrix},
\end{gather}
up to constant terms.

{\bf Second reduction.}  We now perform reduction with respect to the remaining symmetry group, $H/\mathbb{R} \cong \mathbb{R}^2$, using the results from Section~\ref{sec:liegroups}.   We have a left invariant magnetic Lagrangian system on the group $\mathbb{R}^2$, with Lagrangian \eqref{redlagr} and magnetic form $\mathcal{B}_\Gamma = - \Gamma dx \wedge dy$.  The potential $\delta: \mathbb{R}^2 \to \mathbb{R}^2$ corresponding to the latter is given by
\[
	\delta(x, y) = \Gamma \begin{bmatrix}
			-y \\ x
		\end{bmatrix},
\]
and the momentum map is therefore $J_2 (x, y, \dot{x}, \dot{y}) = \mathbb{M}(\dot{x}, \dot{y})^T - \delta(x, y)$.  The non-equivariance 2-cocycle of the momentum map is $\Sigma_\delta = \mathcal{B}_\Gamma$.

The af\/f\/ine action of $\mathbb{R}^2$ on itself is given by $(x, y) \cdot (p_x, p_y) = (p_x - \Gamma y, p_y + \Gamma x)$.  If we f\/ix a~momentum value $(\lambda, \mu) \in \mathbb{R}^2$, the isotropy group $\mathbb{R}^2_{(\lambda, \mu)}$ of the af\/f\/ine action consists of just the zero element, and consequentially the twice-reduced space $J_2^{-1}(\lambda, \mu)/\mathbb{R}^2_{(\lambda, \mu)}$ is nothing but $\mathbb{R}^2$.

The reduced Euler--Lagrange equations~\eqref{twiceredEL} in the case of a left action, assume the following form
\[
	\frac{d}{dt} \begin{bmatrix} p_x \\ p_y \end{bmatrix}
	= - i_{(v_x,v_y)}\Sigma_\delta= \Gamma \begin{bmatrix} -v_y \\ v_x \end{bmatrix},
\]
with $(p_x, p_y)^T = \mathbb{M}(v_x,v_y)^T$, and these are nothing but the equations \eqref{circeqns}.

\subsection*{Acknowledgements}

BL is an honorary postdoctoral researcher at the Department of Mathematics of Ghent University and associate
academic staf\/f at the Department of Mathematics of K.U.Leuven. BL is sponsored by a Research Programme of the Research Foundation -- Flanders (FWO). Part of this work was supported by the Sint-Lucas department of Architecture, K.U.Leuven Association. TM is a  Postdoctoral Fellow of the Research Foundation -- Flanders (FWO).
JV is a postdoc at the Department of Mathematics of UC San Diego, partially supported by NSF CAREER award DMS-1010687 and NSF FRG grant DMS-1065972, and is on leave from a Postdoctoral Fellowship of the Research Foundation--Flanders.
This work is part of the {\sc irses} project {\sc
geomech} (nr.\ 246981) within the 7th European Community Framework Programme.
We are indebted to F.~Cantrijn, M.~Crampin and E.~Garc\'\i a-Tora\~no Andres for many useful discussions. We thank one of the referees for pointing out reference~\cite{mikityuk} on the reduction hypothesis.

%\pdfbookmark[1]{References}{ref}
\LastPageEnding

\end{document}